\documentclass[12pt,showpacs,amsmath,amssymb,aps,floats,aps,floatfix,nofootinbib,12pt]{revtex4}%
\usepackage{epsfig}
\usepackage{dcolumn}
\usepackage{bm}
\usepackage{amsmath}
\usepackage{amsfonts}
\usepackage{amssymb}
\usepackage{graphicx}
\usepackage{latexsym}
\usepackage{rotating}
\usepackage{multirow}
\usepackage{slashed}
\usepackage{xcolor}
\usepackage{hyperref}

\def\nnd{\end{document}}

\def\be{\begin{equation}}
\def\ee{\end{equation}}

\newcommand{\bea}{\begin{eqnarray}}
\newcommand{\eea}{\end{eqnarray}}
\newcommand{\bwt}{\begin{widetext}}
\newcommand{\ewt}{\end{widetext}}

\def\u
\def\hZ{\widehat Z}

\def\eed{\end{document}}

\def\m_z{m_{\textrm {Z}}}

\renewcommand{\u}{\rm{u}}

\def\be{\beta}

\def\sl#1{#1\!\!\!\!/}
\def\rm#1{\textrm{#1}}

\makeatother
\begin{document}
\title{Natural NMSSM confronting with the LHC7-8}
\author{Taoli Cheng$^1$}
\author{Jinmian Li$^1$}
\author{Tianjun Li$^{1, 2, 3}$}
\author{Qi-Shu Yan$^{4,5}$}
\affiliation{$^1$State Key Laboratory of Theoretical Physics,
      Institute of Theoretical Physics, Chinese Academy of Sciences,
Beijing 100190, P. R. China}
\affiliation{$^2$School of Physical Electronics,
University of Electronic Science and Technology of China,
Chengdu 610054, P. R. China}
\affiliation{$^3$George P. and Cynthia W. Mitchell Institute for 
Fundamental Physics and Astronomy,
Texas A\&M University, College Station, TX 77843, USA}
\affiliation{$^4$School of Physics Sciences, University of Chinese Academy of Sciences, Beijing 100049, China}
\affiliation{$^5$Center for High Energy Physics, Peking University, Beijing 100871, China}
\begin{abstract}
The natural supersymmetry (SUSY) requires that stop, sbottom, and gluino be around one TeV or lighter. By using the direct SUSY search bounds from both ATLAS and CMS Collaborations, we examine the constraints on the natural SUSY in the Next to Minimal Supersymmetric Standard Model (NMSSM). We consider two cases of interpretations for the Higgs boson data: (1) the Standard Model (SM) like Higgs boson is the lightest CP-even Higgs boson; (2) the SM like Higgs boson is the second lightest CP-even Higgs boson. We find that the direct SUSY searches at the LHC impose a strong constraint on the light gluino scenarios, and in both cases the gluino can not be lighter than 1.1 TeV
 with sbottom mass up to 600 GeV and stop mass up to 550 GeV.
\end{abstract}

\maketitle

\newpage
\section{Introduction}

As a leading candidate for new physics at the TeV scale, supersymmetry (SUSY) is strongly motivated by solving the quadratic divergence of the Standard Model (SM) and the gauge hierarchy problem as well as by providing a dark matter candidate and a radiative electroweak symmetry breaking mechanism. Compared with the situation before the start-up of the LHC, the discovery of a Higgs boson \cite{atlashiggs,cmshiggs} and the significant constraints from the direct search of the LHC have driven a drastic paradigm shift in the landscape of low energy supersymmetry \cite{Baer:2012yj,Feng:2013pwa}. We now know that the first two generation squarks must be heavier than 1.5 TeV or higher, though it is too early to claim the death of low energy SUSY.  It should be noted that the models with light third generation squarks and/or light gluino, such as the natural SUSY models \cite{Papucci:2011wy}, non-universal gaugino models \cite{Anderson:1996bg}, and  compressed SUSY models \cite{Martin:2007gf}, can still be consistent 
with experimental data. 

The Higgs boson discovered by both the ATLAS~\cite{atlashiggs} and CMS~\cite{cmshiggs} Collaborations can impose significant constraint on some SUSY models. For example, the naive GMSB and AMSB models 
may not produce a SM-like Higgs boson with mass around 125 GeV~\cite{Draper:2011aa} unless stops are very heavy due to the small trilinear soft $A_t$ term (For solutions, see Refs.~\cite{Kang:2012ra, Craig:2012xp}.). Meanwhile, the constrained Minimal Supersymmetric Standard Model (MSSM) or the minimal supergravity (mSUGRA) model may be plagued with the fine-tuning issue in order to accommodate the Higgs boson mass~\cite{mssmhiggsft} via loop-induced contributions. The so-called fine-tuning issue can be greatly alleviated in the next-to MSSM (NMSSM)~\cite{Antusch:2012gv} by utilizing both the tree-level free parameters and loop contributions~\cite{Kang:2012sy,Cao:2012fz,Cheng:2012pe,Agashe:2012zq,Chang:2005ht}.
 
In addition to alleviate the fine-tuning issue in the MSSM, the NMSSM is also well-motivated by solving the $\mu$-problem in MSSM  (for a review see \cite{Ellwanger:2009dp}). The singlino can ease the tension of experimental data and SUSY models (see \cite{Bi:2012jv} for more detailed discussion) in dark matter search. The NMSSM can be embedded into more fundamental theories, say the F-theory Grand Unified Theories (GUTs). The F-theory GUTs can induced unified boundary conditions to free parameters of the NMSSM and yield interesting low energy phenomenologies, as explored in \cite{Aparicio:2012vk}.

The existence of an extra SM singlet in the NMSSM can lead to a richer Higgs phenomenology when compared with that of the MSSM, there are three CP-even Higgs bosons, and two CP-odd Higgs bosons. It is noticed that both two light CP-even neutral Higgs bosons (labelled as $H_1$ and $H_2$, respectively) can be SM-like. As pointed out in \cite{Gunion:2012gc}, the NMSSM can accommodate Higgs boson data quite well if the Higgs boson data observed by the ATLAS and CMS Collaborations were contributed by two degenerate Higgs bosons which have mass around 126 GeV. Moreover, the NMSSM has a better chance to interpret an extra 98 GeV Higgs boson or an extra 136 GeV Higgs boson hinted by the LEP data \cite{Belanger:2012tt} or the Tevatron data \cite{Belanger:2012sd}. It is well-known that the parameter space of either $H_1$ or $H_2$ being SM-like is different from each other~\footnote{While in the NMSSM, the case that the heaviest CP event Higgs $(H_3)$ might be the discovered Higgs boson has been explored in 
Refs.~\cite{Kang:2013rj, Christensen:2013dra}.}, which motivates us to investigate these two cases separately. 

The null results of SUSY search at the LHC significantly and meaningfully constrain the masses of colored particles of SUSY \cite{webcms,webatlas}. The experimental groups usually present their results in the CMSSM and some simplified models without considering any other physical constraints. Recently, quite a few efforts have been devoted to interpret and the LHC search bounds on both the MSSM~\cite{Bechtle:2012zk,Fowlie:2012im,Beskidt:2012sk,Mahmoudi:2012eh} and the NMSSM \cite{Bi:2012jv,Das:2013ta} with full low energy physical constraints.  In these works, the first two generation squarks have usually been excluded up to about 1.4 $\sim$ 1.5 TeV.  Nevertheless, the bounds on
the third generation squarks are weaker due to their small production rates \cite{Cao:2012rz,Bi:2012jv}. The gluino, which is have already excluded up to 1.4 TeV in some constrained model, can be as light as $\sim 500$ GeV, if the mass spectrum is compressed \cite{Dreiner:2012gx,Bhattacherjee:2012mz}.  All this features 
are consistent with the natural SUSY spectrum, which motivate us to examine various scenarios of the natural SUSY, where the the third generation squarks and gluino may be light.

The light gluino can play important roles in the radiative electroweak symmetry breaking, dark matter 
relic density, and gauge coupling unification at high scale. Light gluino scenarios can be well-motivated 
by Grand Unified Theories (GUTs) and string models, for example,  the intersecting D-brane 
models~\cite{Cvetic:2004ui, Chen:2007px, Chen:2007zu},
the F-theory GUTs~\cite{Li:2010mra, Li:2010xr}, 
the G2-MSSM~\cite{Acharya:2008zi},  the unnatural SUSY~\cite{ArkaniHamed:2012gw,ArkaniHamed:2004fb}, the Split SUSY scenario \cite{Giudice:2004tc,Giudice:2011cg,Giudice:1998xp,Wells:2003tf, Dine:2004dv} as well as the PeV (Split) SUSY scenario \cite{Wells:2004di},  the natural SUSY proposed in \cite{Papucci:2011wy}, the Hidden SUSY scenario \cite{Baer:2011ec}, and  the compressed SUSY scenario \cite{Martin:2007gf}.

Due to its large production rate, the light gluino scenarios have been a focus of phenomenological researches. Its discovery potential at the early LHC runs has been explored in literatures. For example, in the reference \cite{Hewett:2004nw}, the signature of a long lived gluino under the split SUSY has been explored. As shown in reference \cite{Chen:2010kq}, a broad and diverse sample of light gluino scenarios (from 350 GeV to 700 GeV) in minimal and non-minimal supergravity models are proposed. In the reference \cite{Giudice:2010wb}, the scenarios of nearly degenerate gaugino masses are considered. The pair production of light gluino can have multi-top final states \cite{Kane:2011zd} and multi-b final states \cite{Ajaib:2010ne}, and it is expected that the multi-lepton and multi-b jet channels are sensitive to light gluino mass region due to the clean SM background.

The lightest supersymmetric particle (LSP) like neutralino can be a cold dark matter candidate.
The light gluino scenarios can also address the dark matter relic density in our Universe via 
the gluino-LSP coannihilation, which motivate us to examine such scenarios with current LHC data. Among them as given in Ref.~\cite{lightgluino1,Chen:2010kq}, the gluino-LSP coannihilation region is represented by points LG3-5. Such a scenario might lead to a long lifetime gluino and chargino, which can have interesting LHC phenomenologies, like the displaced kink appeared in the detectors, as explored in reference \cite{Kane:2012aa} where the signature of gluino decaying to wino-like LSP is considered. The long lived charged wino can also  be captured by the silicon tracker detectors at the ATLAS and CMS experiments. More recent work on the detection of gluino-LSP coannihilaiton can be found in Ref.~\cite{Dreiner:2012gx}, where the search for compressed SUSY by using monojet signature has been carefully evaluated.

Light gluino scenarios accompanied with a light third generation of squarks have been the intense search focus at the LHC. According to the SUSY search results from the ATLAS~\cite{webatlas} and CMS~\cite{webcms} Collaborations, gluino mass has been excluded up to $\sim$ 1.3 TeV with $m_{\tilde q} \simeq m_{\tilde g}$ in the CMSSM/mSUGRA. The upper limit reduces to $\sim$ 750 GeV with decoupled squarks. If interpreted in simplified models, different decay patterns have been considered separately. For the 100\% decay chain: $\tilde g \to t \bar t \tilde \chi^0_1$ (mediated by a virtual $\tilde t$), the allowed gluino mass has been pushed up to $\sim$ 1.2 TeV for the LSP lighter than $\sim$ 400 GeV, and $\sim$ 750 GeV for all available LSP masses. As for the decay chain $\tilde g \to b \bar b \tilde \chi^0_1$ (mediated by a virtual $\tilde b$), $m_{\tilde g} \lesssim 1.2 ~\rm{TeV}$ has been ruled out with $m_{\tilde \chi^0_1} < 500 ~\rm{GeV}$.  A gluino lighter than $\sim$ 1 TeV with $m_{\tilde \chi^0_1} \lesssim 400~\rm{GeV}$ are excluded in the non-b tagging analysis. Bounds for a long-lived $\tilde g$ which can form a R-hadron are also available, where $m_{\tilde g} \lesssim 1~\rm{TeV}$ are excluded by using the signature of slow-moving  objects (low $\beta$, $\beta \gamma$) in the detectors.

It should be noticed that all experimental bounds at the LHC are obtained by using some simple assumptions, where typically the decay branching fraction is oversimplified to be either vanishing or $100\%$. In reality, to evaluate whether these light gluino models are still alive or dead, model dependence must be carefully examined. For such a purpose, a more reliable approach is the Monte Carlo simulation, where model dependence can be correctly accounted for.

In this work, we focus on the bounds to light gluino scenarios in a concrete model---the NMSSM. To incorporate the direct search for SUSY, we assume that squarks of the first two generations are heavier than $1.5$ TeV, while allow the squarks of the third generation and gluino to be light. We explore two cases of interpretations for the Higgs boson data: (1) the lightest CP-even Higgs boson is around $125\sim127$ GeV; (2) the second lightest CP-even Higgs boson is around  $125\sim127$ GeV. 

We find that the gluino-LSP coannihilation region in the natural SUSY models has been ruled out, and the gluino mass must be larger than 1.1 TeV. The reason can be attributed to the fact that the coannihilation region demands that the gluino is around 300 GeV. But, due to its huge cross section, the bounds derived from the associated mono-jet process $pp\to j + {\tilde g} {\tilde g}$ and the $\alpha_T$ analysis approach 
require that the gluino mass be larger than 400 GeV.
It is interesting to note that the $\alpha_T$ analysis approach is also sensitive to the coannihilation region due to a relatively large fraction of boosted data sample in the signal events. Therefore, in the natural NMSSM, in order to be consistent with all experimental data, the current bounds on gluino mass must be larger than 1100 GeV (with sbottom mass up to 600 GeV and stop mass up to 550 GeV).

The paper is organized as follows. In Section II, we describe the Markov Chain Monte Carlo method and our scanning strategy as well as the experimental bounds except the direct SUSY search bounds. In Section III, we tabulate the main direct SUSY search bounds considered in this work, describe our workflow, and present our main numerical analysis. In Section IV, we examine the constraints on the benchmark points proposed in literatures and propose a few new benchmark points for future LHC runs and future colliders. We end this paper with discussions and conclusion.

\section{Scanning strategy}
\subsection{The setup for scanning and the MCMC method}
To be general, we scan the parameter space of the natural NMSSM defined at the electroweak scale. The null search results of the signature of colored squarks at the LHC constrain the first two generation squark masses to be heavier than 1.5 TeV, which motivates us to set their mass parameters as follows:
\begin{equation}
M_{\tilde{Q_{1,2}}}=M_{\tilde{U_{1,2}}}=M_{\tilde{D_{1,2}}}=1.5\,\, \textrm{TeV}
\end{equation}
Then we are left with a 15 dimensional parameter space (For the sake of simplicity, we fix $A_E =0$.) to be considered . To capture the typical features of natural SUSY and light gluino scenarios, we choose the range of these parameters as:
\begin{align}  \label{input}
& 0<\lambda<0.7, 0<\kappa <0.7, 1.1<\tan \beta <30, 100\,\, \textrm{GeV}< \mu <800\,\, \textrm{GeV}, \nonumber \\
& |A_{\lambda}|< 3\,\, \textrm{TeV}, |A_{\kappa}| < 500 \,\,\textrm{GeV}, 100 \,\, \textrm{GeV} < m_{Q_3}, m_{U_3} < 700\,\, \textrm{GeV},\nonumber \\
& 100\,\, \textrm{GeV} < m_{D_3} < 1000 \,\,\textrm{GeV}, |A_t|<5 \,\,\textrm{TeV}, |A_b|<3\,\, \textrm{TeV}, \nonumber \\
& 10\,\, \textrm{GeV} < M_1 < 1 \,\,\textrm{TeV}, 100 \,\,\textrm{GeV} < M_2 < 1 \,\,\textrm{TeV}, 200 \,\,\textrm{GeV} < M_3 < 1.3 \,\,\textrm{TeV} \nonumber \\
&100\,\, \textrm{GeV} < M_{\tilde{l}} = M_{\tilde{E}} < 500 \,\,\textrm{GeV}.
\end{align}
Obviously, it is very difficult to find the typical features of such parameter space since its dimensionality is too high to scan in a grid method or in a random scanning method. Supposed we need 10 points in one dimensional parameter space, then we need $10^{15}$ points at least for the current situation. To circumvent such an issue of computational cost, instead, we adopt the Markov Chain Monte Carlo (MCMC) method~\cite{mcmcbook} in our scanning. 

The MCMC method is a sampling method to generate a chain of points of the parameter space with a density distribution consistent with experimental constraints. In this method, the computational time at its best is proportional to the dimensionality in a linear way, in contrast to the power way in the random scanning method. The MCMC method has been widely adopted in numerical analysis of many research fields, especially in astrophysics \cite{Trotta:2008qt}. Recently, this method has also been applied in SUSY model scanning \cite{Baltz:2004aw, Allanach:2005kz, Allanach:2006jc, de Austri:2006pe}. 

The method is inspired from the Bayesian theorem which can be put as below:
\begin{equation}
p(H|d)=\frac{p(d|H) p(H)}{p(d)}
\end{equation}
where, $p(H|d)$ is defined as the {\it posterior probability} of the hypothesis after taking into account experimental data. $p(d|H)$ is the {\it sampling distribution} of the data assuming the hypothesis is true. And if it is considered as a function of the hypothesis for fixed data, it is called the {\it likelihood function}. $p(H)$ is the {\it prior probability} which represents our state of knowledge before seeing the data. $p(d)$ is the {\it marginal likelihood} which is just a normalization factor in our case and will be ignored from now. 

In this work, $d$ and $H$ denote a set of computed experimental observables in the model and a set of model parameters given in Eq. (\ref{input}), respectively. $p(H|d)$ is the desired distribution in the parameter space of our scanning after taking into account all experimental constraints. $p(H)$ is taken as flat distribution in this work, which is defined as:
\begin{equation}
p(H)=
\begin{cases}
1 & d_{min} < d < d_{max}\,\\
0          &  \textrm{otherwise}\,.\\
\end{cases}
\end{equation}
$p(d|H)$ denotes the likelihood function determined by experimental constraints, which is defined as \begin{equation}
p(d|H)=\prod_i \,\,p(d_i|H)\,,
\end{equation}
where $p(d_i|H)$ denotes the likelihood function of each of experimental constraints.

Main experimental constraints considered in this work are listed in Table~\ref{constraints}. When the Xenon100 results are applied in our scanning, the proton-DM scattering cross section is rescaled by the formula $\sigma_p^{SI} \times \Omega h^2/0.11$. And the likelihood functions $p(d_i|H)$ adopted in this work can be classified into three categories:
\begin{itemize}
\item Exclusion bounds imposed by setting the likelihood to be zero if the point is already excluded and to be one if it is not excluded. For example, to impose a likelihood value for all theoretical points in our scanning, we assign a zero value to a point if it is unphysical or theoretically unacceptable. Here a point is unphysical or theoretically unacceptable have quite a few meanings: the point might run into a Landau pole for some of its physics parameters at some energy scales, or might lead to a unphysical global minimum at the electroweak symmetry scale, or might have a spectra with a tachyonic mass for a particle, or might have a spectra of which the lightest neutralino is not LSP, or might fail to reach a convergent RGE solution, or might possess no electroweak symmetry breaking. And the bound of the Xenon100 and the bounds from Tevatron and LEP on the masses of sparticles and Higgs boson are realized in this way.

\item Upper bounds described by step functions:
\begin{equation}
p(d_i|H)=\frac{1}{1+\exp[\frac{d_i[H]-d_{\textrm{upper}}}{0.01 \,\,d_{\textrm{upper}}}]}\,,
\end{equation}
where $d_i[H]$ means the observable computed in the NMSSM. For example, the bound of dark matter relics density is realized by a step function. Although we also use a step function to the rare decay $B_s \to \mu^+ \mu^-$, we notice that the LHCb collaboration claimed a discovery of this mode with measured branching fraction $3.2^{+1.5}_{-1.2} \times 10^{-9}$ \cite{Aaij:2012nna}. Nonetheless, our main results are not sensitive to this bound. 
\item Physics constraints described by Gaussian functions with well measured central values and deviations
\begin{equation}
p(d_i|H)=\exp[- \frac{(d_i[H]-d_{\textrm{cen}})^2}{\sigma^2}].
\end{equation}
For example, the likelihood function of Higgs boson mass is taken as a Gaussian function with a central value 125 GeV and an allowed deviation is taken as 2 GeV.
\end{itemize}

\begin{table}[htb]
\begin{center}
\begin{tabular}{|c|c|c|c|} \hline
Experimental obserbables &  Mean vaule  &  Deviation  & Ref. \\ \hline
$BR(B^+ \to \tau^+ \nu_{\tau})$ & $1.67 \times 10^{-4}$ & $0.4 \times 10^{-4}$ & \cite{Asner:2010qj} \\
$BR(B \to X_s \gamma)$ &  $3.52 \times 10^{-4}$ & $0.3 \times 10^{-4}$ & \cite{Barberio:2006bi}\\ \hline
\multicolumn{3}{|c|}{$BR(B_s \to \mu^+ \mu^-)$ $< 4.5 \times 10^{-9}$}& \cite{Aaij:2012ac} \\\hline \hline
\multicolumn{3}{|c|}{$\Omega h^2 $ $< 0.136$} & \cite{Larson:2010gs} \\\hline
\multicolumn{3}{|c|}{Xenon100 (2012)} & \cite{Baudis:2012zs} \\\hline \hline
$m_{Higgs}$ & 125 GeV & 2 GeV & \\
$R_{\gamma \gamma}$ & 1.6 & 0.4 & \cite{:2012gu,:2012gk}\\
$R_{VV}$ & 1.0 & 0.2 & \\ \hline
\end{tabular}
\end{center}
\caption{Physical bounds which have been taken into account in our scanning are listed here.}
\label{constraints}
\end{table}

Once the likelihood function, which has incorporated all experimental constraints appropriately, has been specified, we can construct Markov chains through the Metropolis algorithm. The chain is a set of points in our parameter space, which can be labelled as $\{P_0, P_1, P_2, \cdots P_i, P_{i+1} \cdots \}$. We start with a seed (labelled as $P_0$). The chain can be generated by the following steps: 1) For any a point $P_i,\,\,i=0,1,2,\cdots$, we compute its value of likelihood function $P(d|P_i)$. 2) A proposed point in our parameter space labelled as $P_p$ is introduced and its value of the likelihood function is evaluated as $P(d|P_p)$. If $P(d|P_p) > P(d|P_i)$,  this walk is accepted and a new start point is found $P_{i+1}=P_p$. If $P(d|P_p) < P(d|P_i)$, this walk is accepted with a probability $P(d|P_p)/P(d|P_i)$. When this proposed step is accepted, label it as $P_{i+1}=P_p$; when this proposed step is not accepted, the old point $P_i$ will be used, i.e. $P_{i+1}=P_i$. 3) Repeat these two steps, after a 
long and sufficient walk (proportional to the dimensionality of parameter space, say 2.5 million for each case), a Markov Chain with sample points reflected the experimental constraints can be constructed.

As observed in \cite{Baltz:2004aw}, to reduce the computational time while maintaining the sufficient sample points which capture the features of the constrained parameter space can be balanced by utilizing an appropriate proposal step in each a walk. According to the rule of thumb, a step with an acceptance rate around $25\%\sim30\%$ is the best one, which is realized in our scanning by trial and error.

\subsection{Features of the Sampled Points}
\begin{figure}[!htb]
\begin{center}
\includegraphics[width=0.48\columnwidth]{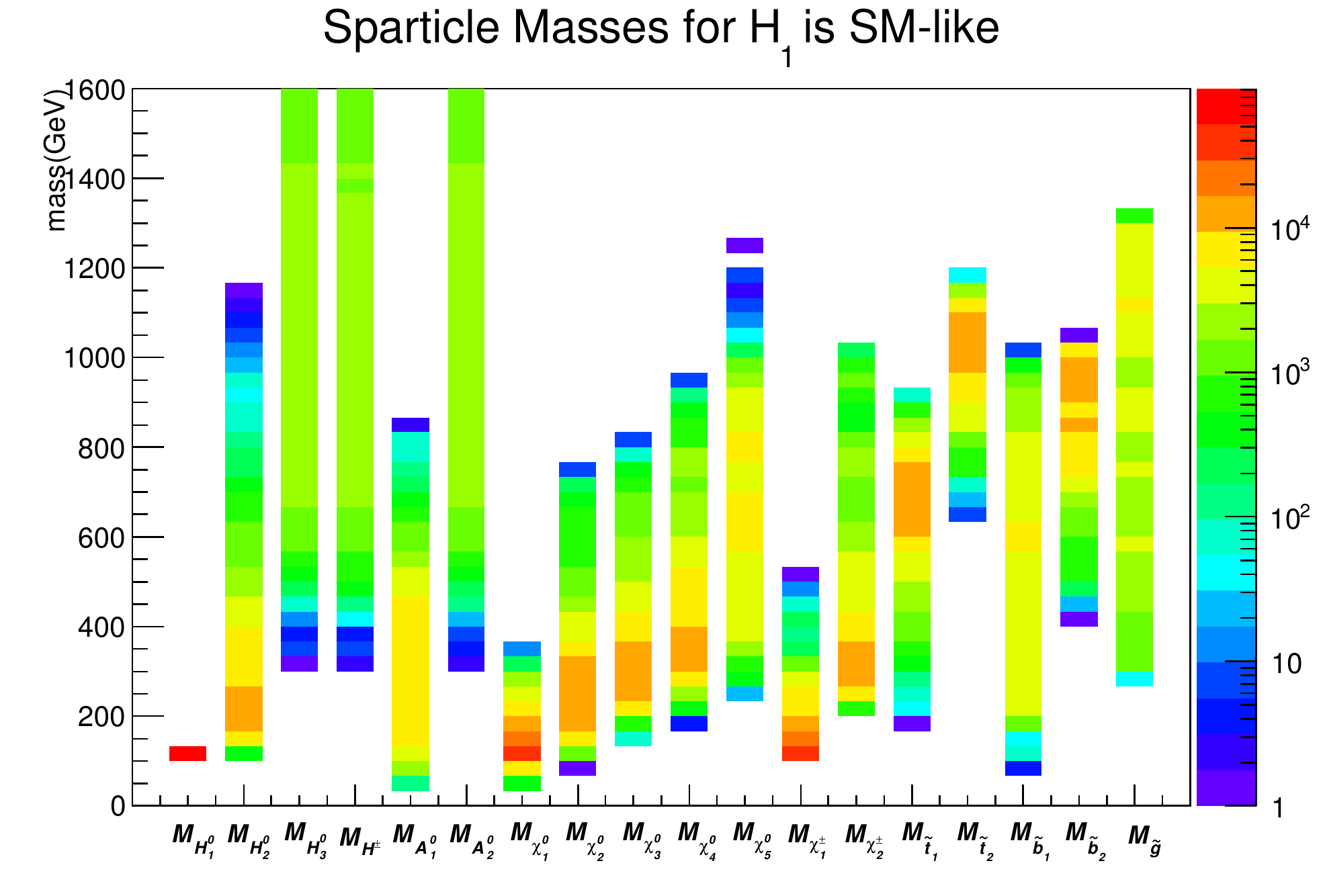}
\includegraphics[width=0.48\columnwidth]{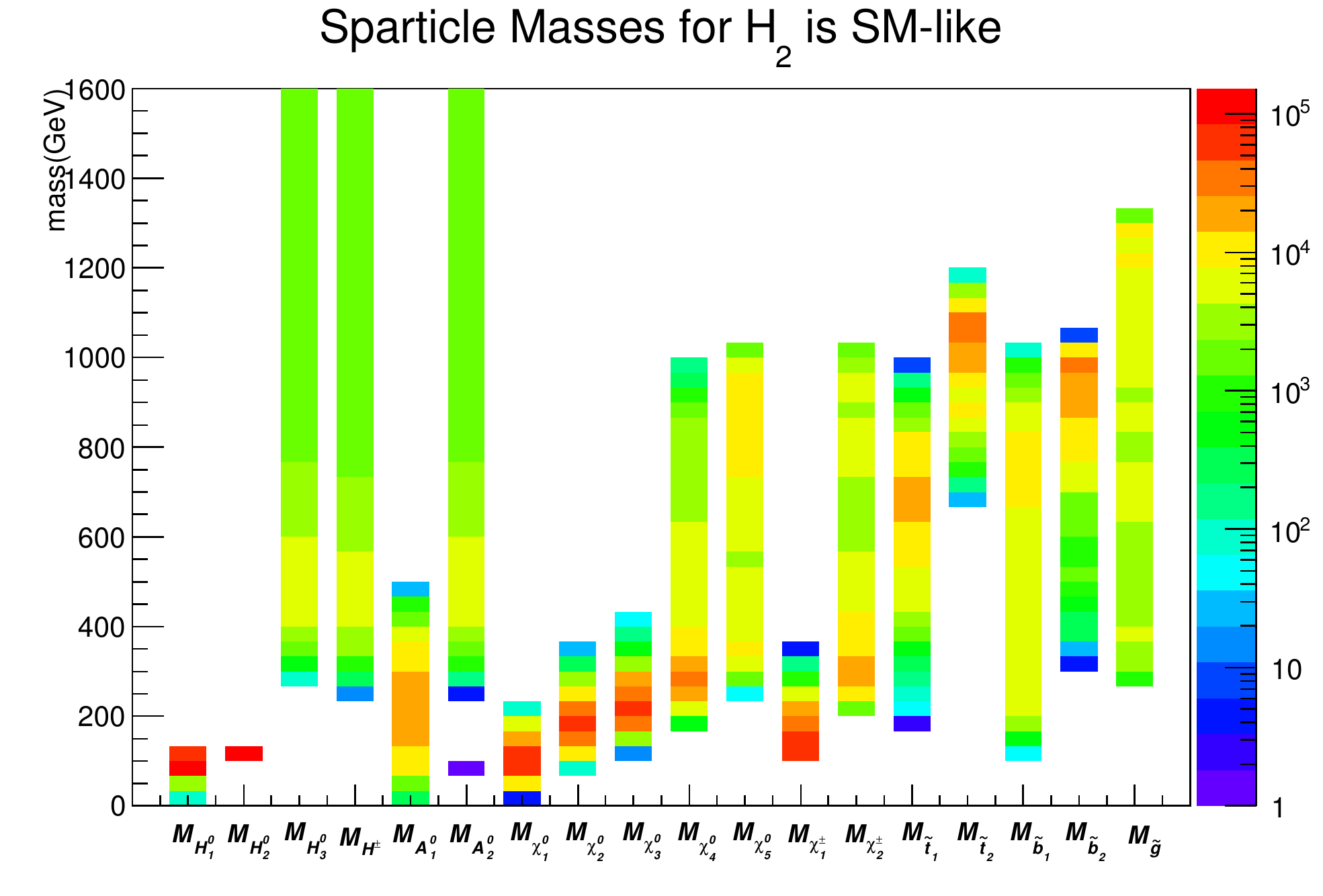}
\caption{The distribution of mass spectra of sparticles are shown for the first and second cases.
\label{massspectra}}
\end{center}
\end{figure}

In our scanning, we consider two cases of the interpretations for the Higgs boson data: in the first case, we assume that the $H_1$ is the SM-like Higgs boson; in the second case, we assume that the $H_2$ is the SM-like Higgs boson. 
We implement the MCMC method in NMSSMtools 3.2.1 \cite{Ellwanger:2004xm,Ellwanger:2005dv} and construct a chain with a 2.5 million points for each case. The distribution of mass spectra of sparticles for both the 1st case and the 2nd case is shown in Fig. \ref{massspectra}.

\begin{figure}[!htb]
\begin{center}
\includegraphics[width=0.49\columnwidth]{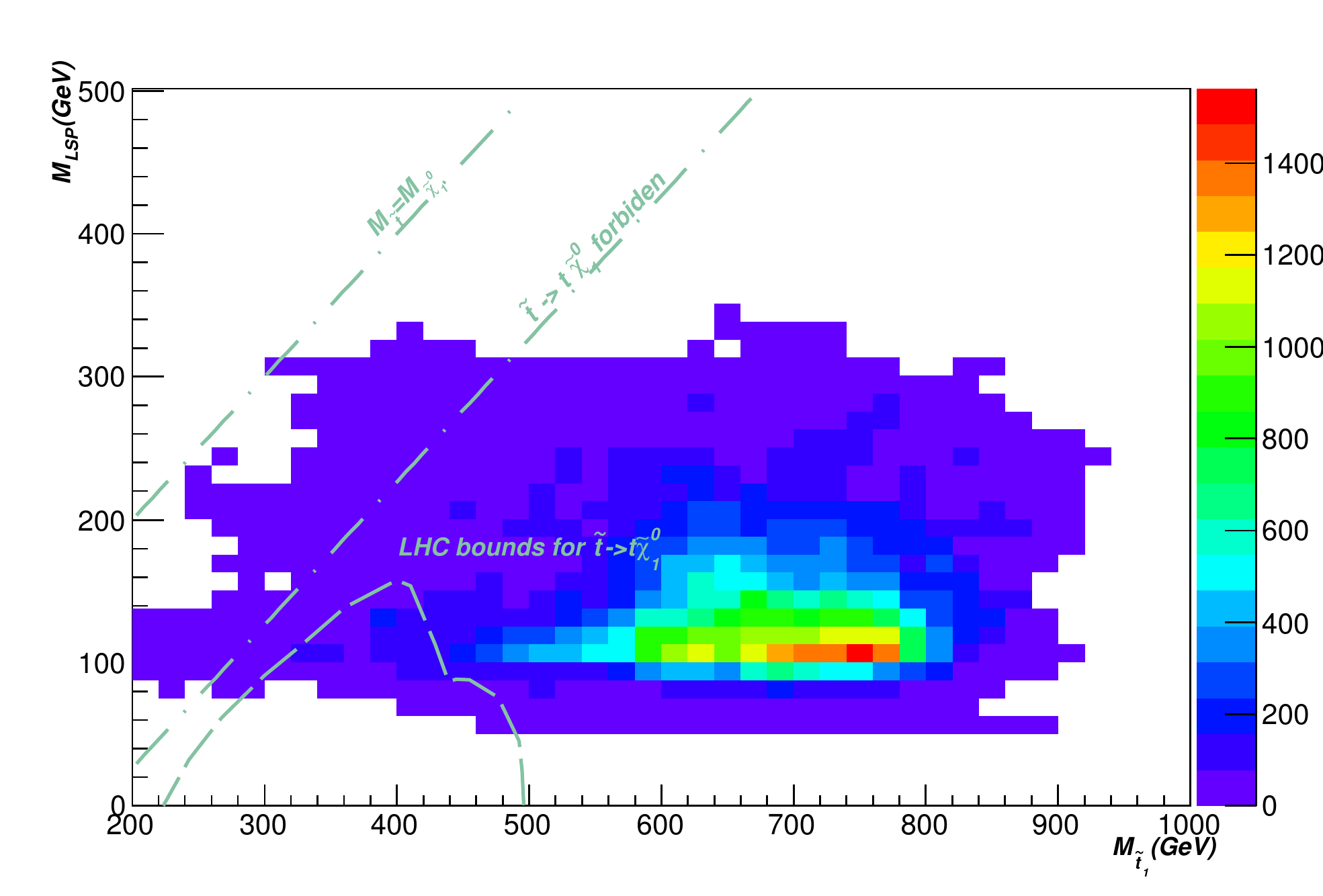}
\includegraphics[width=0.49\columnwidth]{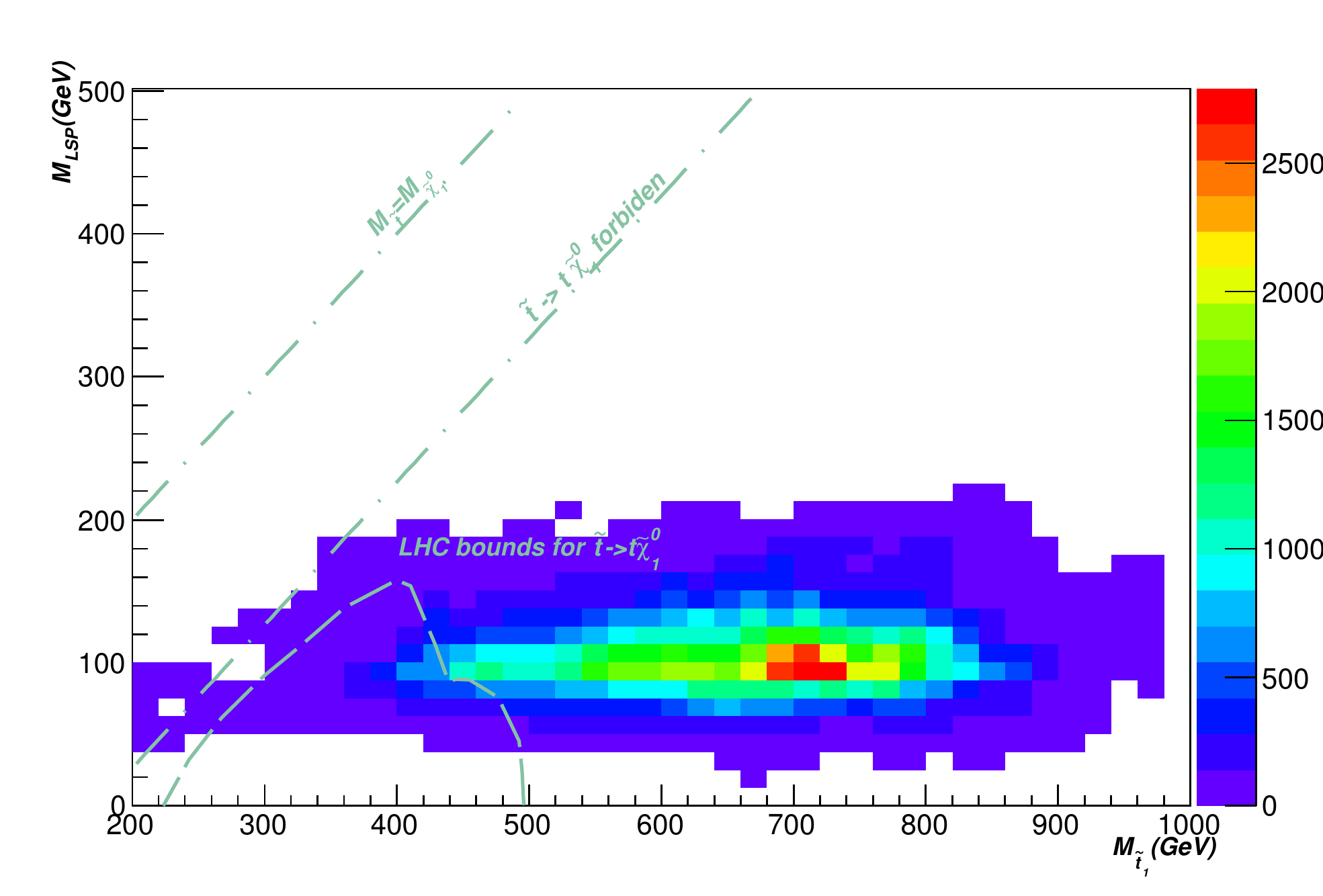}
\caption{The distribution of points in  the $m_{\tilde t_1} - m_{\tilde \chi^0_1}$ plane are shown for the first and second cases.
\label{stop-lsp}}
\end{center}
\end{figure}

\begin{figure}[!htb]
\begin{center}
\includegraphics[width=0.49\columnwidth]{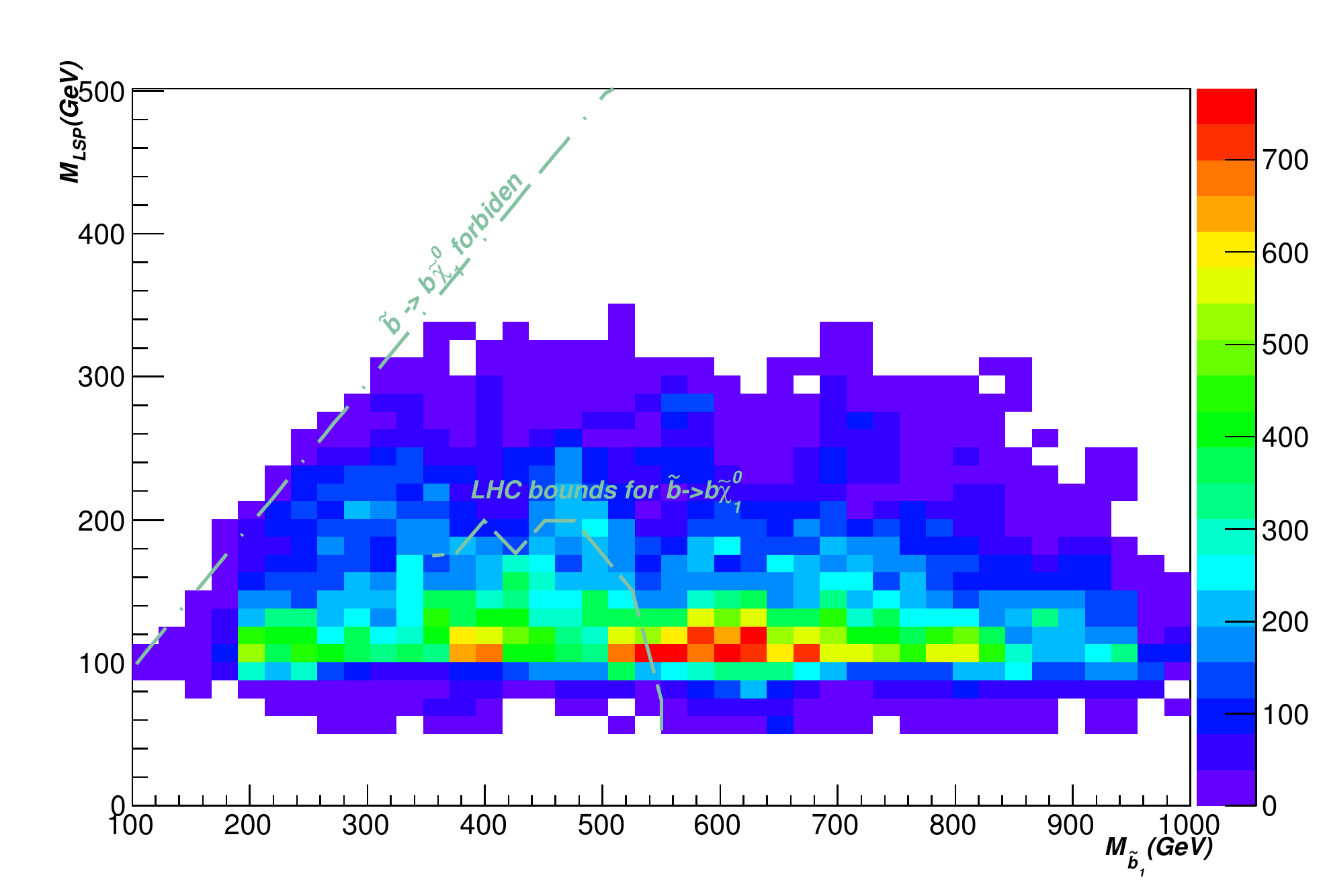}
\includegraphics[width=0.49\columnwidth]{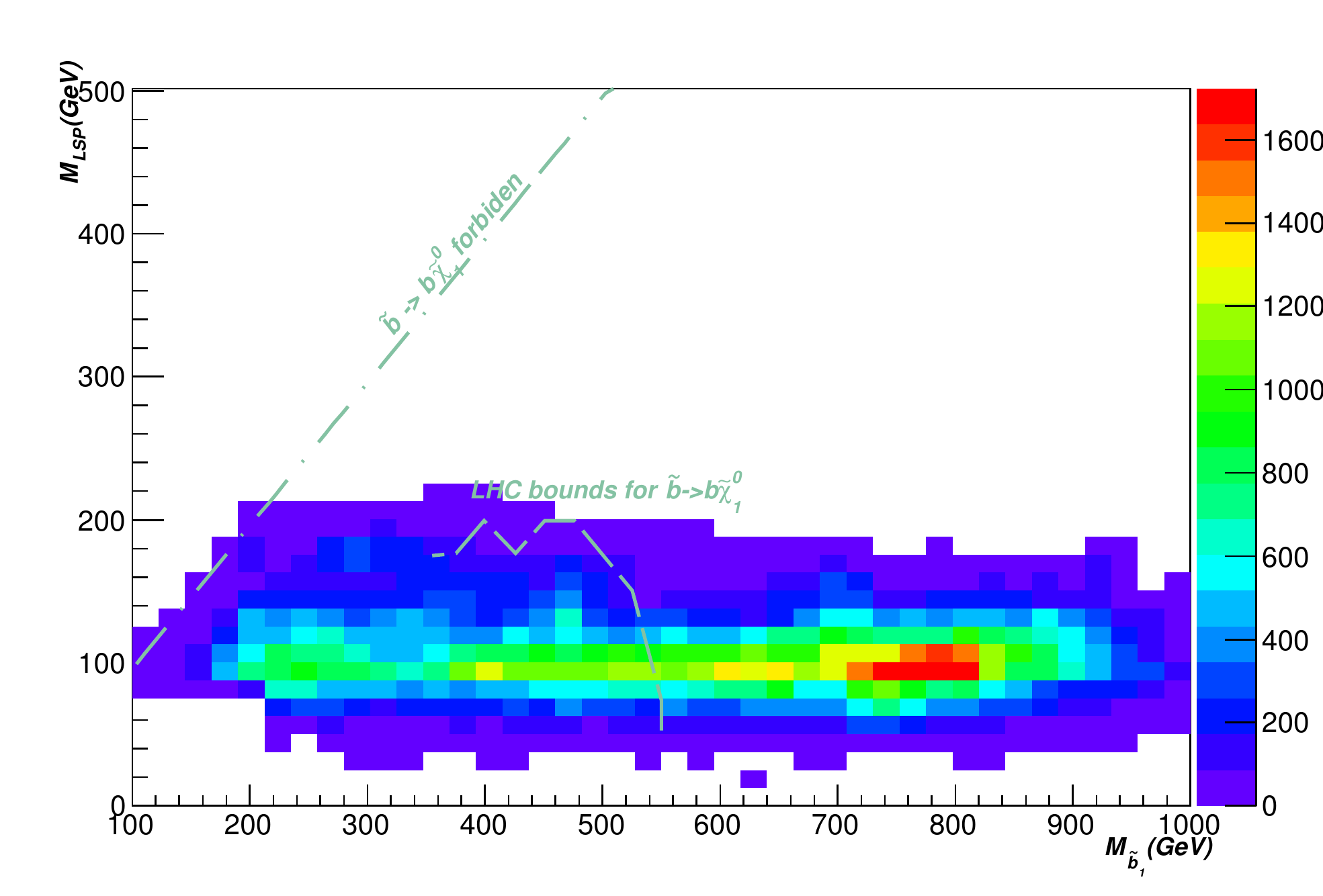}
\caption{The distribution of points in  the $m_{\tilde b_1} - m_{\tilde \chi^0_1}$ plane are shown for the first and second cases.
\label{sbot-lsp}}
\end{center}
\end{figure}

\begin{figure}[!htb]
\begin{center}
\includegraphics[width=0.49\columnwidth]{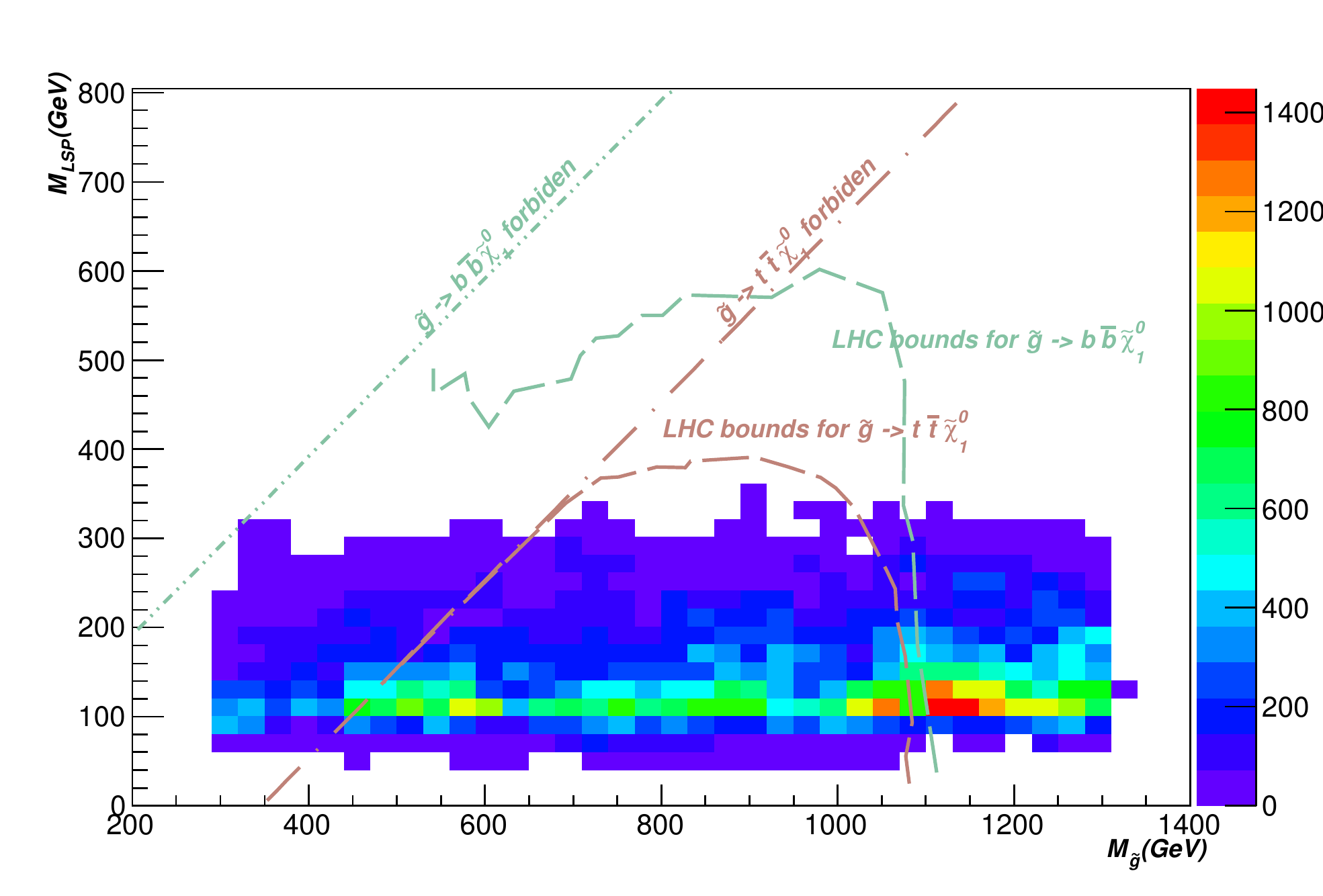}
\includegraphics[width=0.49\columnwidth]{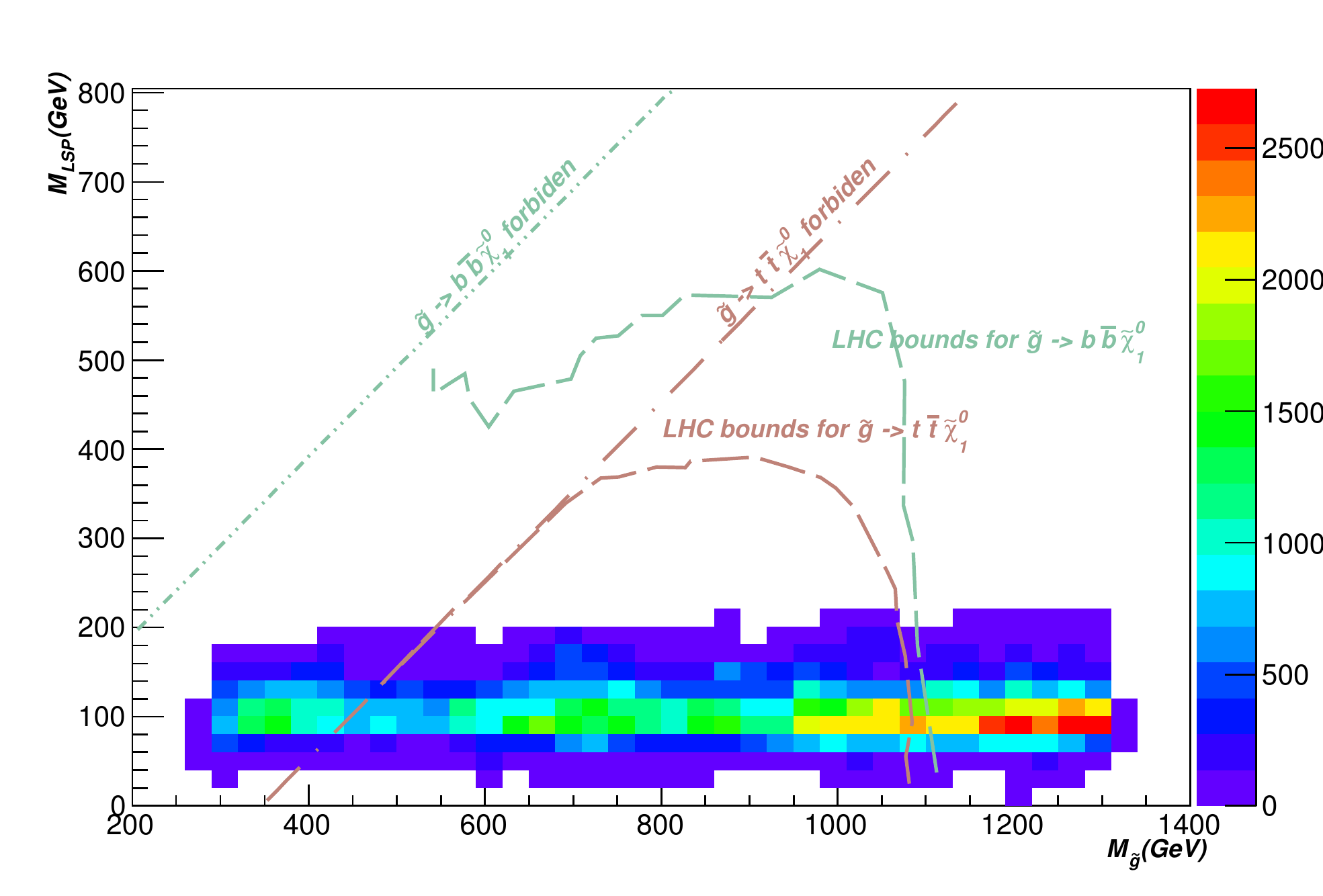}
\caption{The distribution of points in  the $m_{\tilde g_1} - m_{\tilde \chi^0_1}$ plane are shown for the first and second cases.
\label{gluino-lsp}}
\end{center}
\end{figure}

There are a few comments on Fig. (\ref{massspectra}) in order: 
\begin{itemize}
\item For the Higgs sector, in the 1st case, the mass of $H_2$ can spread in a quite large range from 120 GeV up to 1200 GeV. Similarly $H_3$, $A_2$ and $H^\pm$ are quite heavy and also expand in a large range from 300 GeV up to 1600 GeV. In contrast, in the 2nd case,
the mass of $H_1$ is confined to be smaller than $H_2$, therefore quite a fraction of $H_3$, $A_2$ and $H^\pm$ situate below 1000 GeV. 
\item For the neutralino sector, in the first case, the lightest neutralino can spread from a few GeV to 340 GeV, and most of them are smaller than 200 GeV. The second and third neutralinos can expand from a few ten GeV to 700 GeV. In contrast, in the 2nd case, the LSP is compressed in a much smaller mass range from a few GeV to 220 GeV, while most of them situate near 100 GeV. The second and third neutralinos are also compressed in much smaller ranges.
\item For the chargino sector, in the first case, the lighter chargino can expand from 100 GeV to 500 GeV. In contrast, the lighter chargino can only expand from 100 GeV to 300 GeV  in the second case.
\item It is interesting to noticed that the distribution of stop squarks, sbottom squarks and gluino is similar in both cases. It is remarkable that in the second case, the gluino-LSP coannihilation region is not allowed, while for the first case, such region is possible.
\end{itemize}
We also show the features of these points when projected on the $m_{\tilde t_1} - m_{\tilde \chi^0_1}$ plane, the $m_{\tilde b_1} - m_{\tilde \chi^0_1}$ plane, and the $m_{\tilde g} - m_{\tilde \chi^0_1}$ plane, as shown in Figs. (\ref{stop-lsp}-\ref{gluino-lsp}). One general feature shown in Figs. (\ref{stop-lsp}-\ref{gluino-lsp}) is that the favored mass of LSP is around 100 GeV for both cases. In the first case, the LSP is Bino or Wino dominant and its mass is determined by parameters $M_1$ and $M_2$, which is similar to the case of the MSSM. In the second case, the LSP is either higgsino or singlino dominant and its mass range is determined by $\mu$, $\lambda$, $\kappa$, etc. 

From Fig. (\ref{stop-lsp}), we can read that both cases favor a relative heavy ${\tilde t_1}$ (say around 750 GeV in the first case and 700 GeV in the second case), such a tendency is determined by the Higgs boson mass. 

It is remarkable that the second case allows a narrower LSP mass range than the first case. Such a feature can be understood by a correlation existed between the singlet scalar mass and the singlino mass. The mass formulae of them are provided below:
\begin{equation}
\label{lspmass}
M_{H,S}^2=\lambda^2 v^2 A_{\lambda} \frac{\sin 2 \beta}{2 \mu} +4 \kappa^2 s^2+A_{\kappa} \kappa s\,\,,
\end{equation}
\begin{equation}
M_{\tilde S}=2 \kappa s\,\,.
\end{equation}

In the second case, it is required that the lightest Higgs boson be mainly singlet-like, which will set an upper limit for $M_{H,S} \lesssim 125 $ GeV. To guarantee the second Higgs boson mass to be $m_{H_2} \sim 125$ GeV, a cancellation condition given below
\begin{equation}
1-(A_{\lambda}/2\mu+\kappa/\lambda) \sin 2 \beta \simeq 0 \,,
\end{equation}
is demanded, as pointed out in \cite{Kang:2012sy}. With this condition, from Eq. (\ref{lspmass}) we obtain an inequality 
\begin{equation}
4 \kappa^2 s^2 +A_\kappa \kappa s + \lambda^2 v^2 (1-\frac{\kappa \sin 2 \beta}{\lambda}) < 125^2 \,.
\end{equation}
From this inequality we can solve out an $s_{\textrm{max}}$ which is given as 
\begin{equation}
\kappa s_{max}=\frac{1}{8}( |A_\kappa|+\sqrt{500^2+A_\kappa^2-16\lambda^2 v^2 (1-\frac{\kappa \sin 2 \beta}{\lambda}) })\,,
\end{equation}
since  $(1-\frac{\kappa \sin 2 \beta}{\lambda}) > 0$  and most of $|A_\kappa|$ can not be larger than $300$ GeV in our scanning (after imposing all experimental cuts), $\kappa s_{max}$ should be smaller than $110$ GeV. The only exception occurs when  $M_{H,S} \lesssim 125 $ GeV due to the large mixing between the singlet and the doublet $H_d$. Then the singlet can have a mass $M_{H,S} > 125 $ GeV and become less than $125$ GeV after a large mixing with $H_d$. However, this kind of space needs a certain degree of fine-tuning.  

Moreover, such a feature can also be understood in that a relatively small $\mu_{eff}$ ($\sim 100 - 200 \rm{GeV}$) are favored to produce an appropriate S-$\rm H$ mixing so as to yield the right SM-like Higgs boson mass. Consequently, we have $M_{\tilde s} = 2 \kappa s = 2 \frac{\kappa}{\lambda} \mu \lesssim \mu$, which result in a light LSP within a narrower range. 

In contrast, for the first case, the LSP is Bino or Wino dominant and its mass range is simply determined by the range of parameters $M_1$ and $M_2$ and there is no such a correlation exists. It is noticed that in the first case, the stop-LSP coannihilation region is allowed, while in the second case, such a region is missing.

In Fig. (\ref{sbot-lsp}), it is obvious that the first case allows more points for the sbottom-neutralino coannihilation regions. But generally speaking, the Higgs boson mass does not affect the distribution of mass of ${\tilde b_1}$, since the contribution of ${\tilde b_1}$ to the Higgs boson mass is much smaller when compared with that of ${\tilde t_1}$ when $\tan\beta < 30$. 

In Fig. (\ref{gluino-lsp}), it is noticed that since the gluino can contribute to the Higgs boson mass via two loops, the mass of gluino can be affected by the Higgs boson data. The most favored gluino mass is around 1.1 TeV and 1.2 TeV for the two cases, respectively. 

To appreciate how stringent the LHC bounds can reach, in Figs. (\ref{stop-lsp}-\ref{gluino-lsp}), we deliberately show the combined LHC bounds derived from the simplified models. In these Figs., we combined the available bounds from both ATLAS and CMS collaborations \cite{lhcbounds}. In Fig. (\ref{stop-lsp}), we use the bounds of $pp\to {\tilde t} {\tilde t}$ with $\tilde t \to t \tilde \chi_1^0$ from the ATLAS analysis based on the dataset of 4.7 $\rm{fb}^{-1}$ and $\sqrt{s}=7$ TeV. In Fig. (\ref{sbot-lsp}), we adopt the bounds of $pp\to {\tilde b} {\tilde b}$ with $\tilde b \to b \tilde \chi_1^0$ from the CMS analysis based on the dataset of 4.98 $\rm{fb}^{-1}$ and $\sqrt{s}=7$ TeV by using the $\alpha_T$ variable. In Fig. (\ref{gluino-lsp}), we compile two types of bounds: 1) $pp\to {\tilde g} {\tilde g}$ with $\tilde g \to t \bar t \tilde \chi_1^0$ form the ATLAS collaborations based on the dataset of 5.8 $\rm{fb}^{-1}$ and $\sqrt{s}=8$ TeV and 2) $\tilde g \to b \bar b \tilde \chi_1^0$ from the CMS 
collaborations based on the dataset of 4.98 $\rm{fb}^{-1}$ and $\sqrt{s}=7$ TeV.

One obvious concern is whether all those allowed points within the bound curves are still alive or whether all points outside the bound curves are safe. To address this question, we sample 2400 points from these 5 million points (including both two cases of interpretations of Higgs boson data) to perform an analysis of the constraints from the direct SUSY searches by the LHC experiments. We have examined that the distributions of these 2400 points are similar to those 5 million points, therefore the analysis of these 2400 points can be reliably extrapolated to the sample of 5 million points. 

\section{Constraints from Direct SUSY searches at the LHC}
\subsection{SUSY search bounds from LHC experiments}
To study the constraints from the direct SUSY searches by the LHC collaborations, we implement in our analysis both results of ATLAS and CMS collaborations for the datasets accumulated with the collision energy $\sqrt{s}=7$ TeV and $\sqrt{s}=8$ TeV, shown in Table (\ref{susybounds}), where the direct SUSY search channels from both ATLAS and CMS collaborations and the references are tabulated. Below we briefly describe these search channels and their sensitivity to possible SUSY signals:
\begin{table}[th]
\begin{center}
\begin{tabular}
[c]{|c|c|c|}\hline
 channels                             & ATLAS                  				& CMS \\\hline\hline
jets + ${\sl E}_T$                      & \cite{:2012rz,Aad:2011ib,AC11155,AC11086,daCosta:2011qk,AC10065},\cite{AC12109}$^8$ & \cite{CMS:12005} \\\hline
 multi-jets + ${\sl E}_T$                & \cite{Aad:2012hm,Aad:2011qa},\cite{AC12103}$^8$                                       & \cite{:2012mfa} \\\hline \hline
B-Jets + jets + ${\sl E}_T $            &       \cite{:2012si,:2012pq,Aad:2011cw,AC12106,AC11098,AC10079}              &\cite{:2012rg},\cite{CMS:12016}$^8$ \\\hline   
 B-Jets + leptons + jets + ${\sl E}_T$   &      \cite{:2012yr,:2012ar,ATLAS:2012ah,AC11130,Aad:2011ks}                  &\cite{CMS:11028},\cite{CMS:12017}$^8$\\\hline 
letpons + jets +  ${\sl E}_T$           &     \cite{:2012uu,:2012ms,:2012tx,ATLAS:2012ai,Aad:2011cwa,ATLAS:2011ad,AC12140,AC11090,Aad:2011xm,Aad:2011xk,Aad:2011hh,AC11091,AC11064,AC10066},\cite{AC12104,AC12105}$^8$      &\cite{CMS:12010} \\\hline
 multi-letpons + ${\sl E}_T$             &     \cite{:2012ku,:2012gg,:2012cwa,AC12108,AC12035,AC12001,AC11039}      &   \cite{Chatrchyan:2012mea}\\\hline
Z-boson + jets + ${\sl E}_T$         &         &\cite{Chatrchyan:2012qka}\\ \hline
 Mono-jet +  ${\sl E}_T$                           &        \cite{AC12084}   &\cite{Chatrchyan:2012me}\\ \hline             
\end{tabular}
\end{center}
\caption{The direct SUSY search results from ATLAS and CMS collaborations are tabulated, where the upper script $8$ in the channels denotes the results being obtained from the dataset with the collision energy $\sqrt{s}= 8$ TeV. }
\label{susybounds}
\end{table}
\begin{itemize}
\item The jets +${\sl E}_T$ channel is the classical search channel for the signature of pair-production of squarks and gluino. In this search channel, a large $H_T$(defined as the scalar sum of $p_T$ of all reconstructed jets in one event) and a large missing transverse energy (denoted as ${\sl E_T}$) are required. The hard $H_T$ is expected if all jets are produced from the heavy SUSY particles decay.  And a large ${\sl E_T}$ predicted in R-parity conserving SUSY can efficiently suppress QCD background.

Apart from these two simple cuts, some characteristic kinematic variables, such as
the $\alpha_T$ variable, the $M_{T2}$ variable, and the Razor variable, are used to discriminate SUSY signal from the SM background. Below we describe these kinematic variables in order.

As introduced in \cite{Randall:2008rw}, the kinematic variable $\alpha_T$ is designed
to distinguish the real ${\sl E_T}$ from the hard process and the pseudo-${\sl E_T}$ from mis-measurement of jet. It's defined as:
\begin{equation}
  \alpha_T=\frac{E_T^{j_2}}{M_T},~M_T=\sqrt{(\sum \limits_{i=1}^2
E_T^{j_i})^2-(\sum \limits_{i=1}^2 p_x^{j_i})^2-(\sum \limits_{i=1}^2
p_y^{j_i})^2},
\end{equation}
for any a 2-jets final state (a multijet final state can be regrouped into a two-jet final state by 
using the combination algorithm that minimises the $E_T$ difference between the two pseudo-jets \cite{Chatrchyan:2012wa}), where $E_T^{j2}$ denotes the $\sl{E}_T$ of less energetic jet. A ${\sl E_T}$ from SUSY particle decay favors a $\alpha_T$ with values greater than $0.5$, while a 
${\sl E}_T$ from mis-measurement of jet energy typically leads to a $\alpha_T$ with values smaller than $0.5$.

The kink variable $M_{T2}$ is introduced in \cite{Lester:1999tx}, it is supposed to determine the transverse mass of a new particle from its pair production with each particle decaying to a visible daughter and an invisible one. It is expected that two reconstructed transverse masses of each particle in each an event should be the same or close to each other. Similar to the $\alpha_T$ variable, a multijet final state can be regrouped into a two-jet final state by using the hemisphere algorithm. Typically, SUSY signals can have a larger $M_{T2}$ around several hundred GeV, while the background of the SM favor a smaller $M_{T2}$.

The Razor variable is introduced in \cite{Rogan:2010kb}, it is defined by CMS collaboration as:
\begin{equation}
  R \equiv \frac{M_T^R}{M_R} ~,
\end{equation}
where $M_T^R$ and $M_R$ are defined as: 
\begin{equation}
  M_R \equiv  \sqrt{(E_{j1}+E_{j2})^2-(p_z^{j1}+p_z^{j2})^2} ~, 
\end{equation}
\begin{equation}
  M_T^R \equiv \sqrt{\frac{{\sl E_T}(p_T^{j1}+p_T^{j2})-\overrightarrow{\sl
E_T}\cdot(\overrightarrow p_T^{j1}+\overrightarrow p_T^{j2})}{2}} ~,
\end{equation}
respectively. This variable has been used to search for SUSY signals 
with colored sparticles in pair production and decaying into invisible particle and jets. 
Signal events are characterized by a large $M_R$ and a large R (which peaks around
0.5 while QCD multijet background events peaks around zero). In this study, we have not taken into account the bounds obtained from the Razor approach, which will be included in our future work.

\item The multi-jets + ${\sl E}_T$ channel is well-motivated by the signals of light squarks of third generation and signals of gluino decaying to squarks of third generation. Typically, such a signal can lead to more energetic jets, when compared with the signal of squarks of the first two generations. For example, the signal from $p p \to \tilde g \tilde g \to (t \bar
t \tilde \chi_1^0)(t \bar t \tilde \chi_1^0)$ with hadronic top quark decays can yield lots of jets in the final state. This search channel should be sensitive to such type of signal.

\item The B-Jets + jets + ${\sl E}_T$ channel can utilize the b-tagging technique, which can be very powerful to reject QCD background. This search channel can improve the sensitivity to signals of third-family squarks production and signals of gluino pair-production with gluinos decaying to the third generation squarks (both on-shell and off-shell) which can prdocue many b jets in the final states. 

\item The B-Jets + leptons + jets + ${\sl E}_T$ channel can utilize both the b-tagging technique and
the lepton(s) (single one or two same-signed dileptons) and can reliably suppress the huge QCD background. It is supposed to be sensitive to the signal of four top final states from gluino pair production with ${\tilde g} \to t {\bar t} + \sl{E}_T$ and $pp\to {\tilde t} {\tilde t}\to t {\bar t} + \sl{E}_T$.

\item The leptons + jets + ${\sl E}_T$ channel can utilize the high efficiency of lepton identification and significantly reject the QCD background. One single lepton, oppsite signed di-lepton and same singed dilepton channels have been considered by experimental collaborations. The channel is expected to be sensitive to $pp\to {\tilde t_1} {\tilde t_1}$ and multi-top final states.

\item For the multi-letpons + ${\sl E}_T$ search channel, three or more well isolated leptons
are required. Trilepton channel would be golden channel to explore chargino and
neutralino pair-production which decay to LSP and leptons mediated by sleptons. In our scanned parameter regions, we observe that the chargino and neutralino can be very light, as shown in Fig. \ref{massspectra}. Consequently, their production rate can be very large and should be considered.

\item The Z-Boson + Jets + ${\sl E}_T$ search channel is supposed to utilize two isolated leptons from a Z boson decay. Both the momenta, sign and flavor of each leptons can be measured quite well. The Z boson peak can be reliably reconstructed. This channel is designed to explore the topology with Z boson produced through neutralino decaying in the cascade decay of colored sparticles.

\item The mono-jet search channel focuses on one single energetic jet originated from the initial state radiation. Typically, the $p_T$ of jet is required to be larger than $\sim$ 100 GeV. And a large missing energy (${\sl E_T}>200~\textrm{GeV}$) is required. This search channel can be sensitive to those co-annihilation scenarios, where the NLSP is almost degenerate with the LSP.
\end{itemize}

ATLAS collaboration have provided upper limits for new physics in their documents and we use those upper limits directly. While similar upper limits are missing in the documents of CMS collaboration. To extract these upper limits from the CMS collaboration, we use the method proposed in \cite{Read,Cowan:2010js} by assuming there is a $30\%$ uncertainty on the possible new physics signal.

\subsection{SUSY Experimental Bounds Implement}
Below we will outline the main procedure as how to implement the SUSY experimental bounds in our study. 

For each a point selected from the constructed Markov chains, we use the NMSSMtools3.2.1 \cite{Ellwanger:2004xm,Ellwanger:2005dv} to generate its mass spectra and decay table in SLHA format.  The mass spectra are used to evaluate the cross sections of SUSY signals. For all points in our work, the most important processes include $pp \to \tilde{g}\tilde{g}$, $pp\to \tilde{t}_1\tilde{t}_1$, $pp\to \tilde{b}_1\tilde{b}_1$ \footnote{We have taken into account the contribution of the process $pp\to \tilde{b}_2\tilde{b}_2$ and have found that the LHC bounds start to constrain those cases with $M_{\tilde b_2} < 600$ GeV.}, $pp \to \tilde{\chi}_i \tilde{\chi}_j$ (where $\tilde {\chi_i}$ include both neutralinos and charginos) pair production. We notice typically that the cross sections of $pp\to \tilde{\chi}_i^0 \tilde {\chi}^0_j$, $pp\to \tilde{\chi}_i^\pm \tilde {\chi}^0_j$, $pp\to \tilde{\chi}_i^\pm \tilde {\chi}_j^\mp$ can be significantly large due to their small masses, like in the second case.

The NLO cross section is evaluated by using the package prospino2 \cite{Beenakker:1996ed}, which will be used to normalize the number of signal events in our analysis. Then, the mass spectra and decay tables are passed to the package MadGraph5 \cite{Alwall:2007st}, and signal events with upto two additional radiative jets for the processes $pp \to \tilde{g}\tilde{g}$, $pp\to \tilde{t}_1\tilde{t}_1$, $pp\to \tilde{b}_1\tilde{b}_1$, $pp \to {\tilde \chi}_i {\tilde \chi}_j$ are generated. To avoid double counting issue in the matrix element calculation and the parton shower simulation, we adopt the MLM-matching scheme with the variable $xqcut=100 GeV$. Then Pythia6 \cite{Sjostrand:2006za} is used to decay the sparticles to the particles of the SM at parton level and to simulate parton shower and hadronization. We use PGS4 \cite{pgs4} to implement fast detector simulation. To reconstruct jets in the final objects, we adopt the anti-kt jet algorithm with the cone size parameter $R=0.5$, and assume the b-tag 
efficiency to be $60\%$ in accompany with a mis-tagged rate for charm quark jet as $10\%$ 
and for other light quark jet as $1\%$, respectively. 

We generate 50,000 events for each of signal processes at parton level, after matching typically we arrive at 30,000 matched events or more. The matched events will be passed to our SUSY bound analysis package to evaluate how many number of events can survive after implementing all experimental cuts. 

To analyze the bounds imposed by the direct SUSY search at the LHC, we develop a systematic analysis package. The main goal of the package is to implement the SUSY constraints given by ATLAS and CMS collaborations in an ease way. With the help of our package, we can evaluate whether a model is still alive or has been ruled out. Up to now, the searches that we have implemented are given in the Table~\ref{susybounds}, and we are upgrading our package by including new LHC bounds released recently. 

For each search channel by feeding the matched events of signal processes to our package, we can read out the selection efficiency for in each signal region. This selection efficiency finally is  translated into the observed number of signal events in each signal region after cross sections and luminosity are taken into account. In order to perform an analysis similar to the LHC collaborations, for each a point we generate two independent event samples with the collision energy $\sqrt{s}=7$ TeV and $\sqrt{s}=8$ TeV, respectively.

In this work, the exclusion limits up to the observed $95\%$ confidence level for each search channel in each signal region have been applied. Accordingly, we define the ratio $R=\frac{N_{\textrm{signal~number}}}{N_{\textrm{observed~limit}}}$ for each signal region in each searching channel. To derive the most stringent constraint, we choose the maximal value of $R$ among all search channels at all signal regions. Obviously, for a model at a specific search channel and a specific signal region, the ratio $R$ is greater than 1 means it has been ruled out by experiments (although we have not taken into account errors, neither Monte Carlo errors nor the fast detector simulation errors). Our package yields all $R$s of each signal region at each search channel. By comparing the $R$s, we can find out the strongest bound, which is denoted by $R_{\textrm{max}}$.

\subsection{Numerical Analysis}
\begin{figure}[!htb]
\begin{center}
\includegraphics[width=0.7\columnwidth]{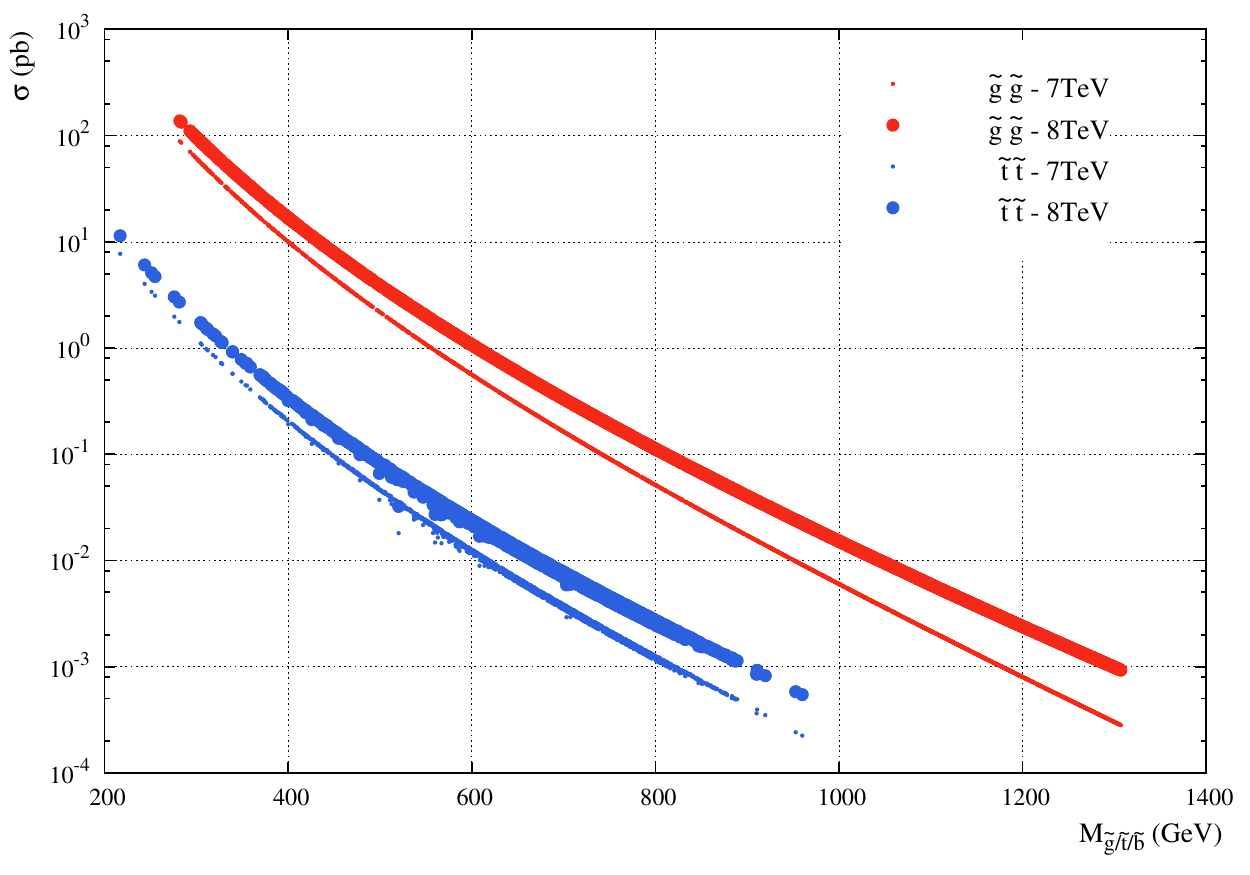}
\caption{The cross sections in varying with the mass of gluino and stop are shown here. The cross section of sbottom is similar to that of the stop and is omitted.
\label{xsec}}
\end{center}
\end{figure}

Here we present our main results of numerical analysis. In Fig. (\ref{xsec}), we present the cross sections of gluino and stop in our scanning. We highlight two observations: 1) Roughly speaking, the cross sections of stop and gluino increase by a factor two when the collision energy increases from 7 TeV to 8 TeV; 2) when stop and glunio have the same mass, the cross section of gluino is 50 times larger than that of stop. It is also noticed that the cross sections are almost equal for the stop pair production and the gluino pair production if the mass of $m_{\tilde t}\approx m_{\tilde g} - 250$ when $m_{\tilde g}=600$ GeV and $m_{\tilde t}\approx m_{\tilde g} - 400$ GeV when $m_{\tilde g}=1200$ GeV. At the tree-level, the cross sections of $pp\to {\tilde t} {\tilde t}$ and $pp\to {\tilde g} {\tilde g}$ are simply determined by the mass parameters, while at the NLO level, colored sparticles at loop can contribute and lead to a minor change in the cross section. It is noticed that the fluctuation in the 
cross section of the process $pp\to {\tilde t} {\tilde t}$ near the mass region $500 \sim 600$ GeV by a few points is originated by the stop-decay threshold ($m_{\tilde t}=m_t+m_{\tilde g}$) effect\cite{Beenakker:2010nq,Beenakker:1997ut}. The cross sections of $pp\to {\tilde b} {\tilde b}$ is similar to that of $pp\to {\tilde t} {\tilde t}$. Therefore we neglect them in the Fig. (\ref{xsec}).

\begin{figure}[!htb]
\begin{center}
\includegraphics[width=0.32\columnwidth,angle=270]{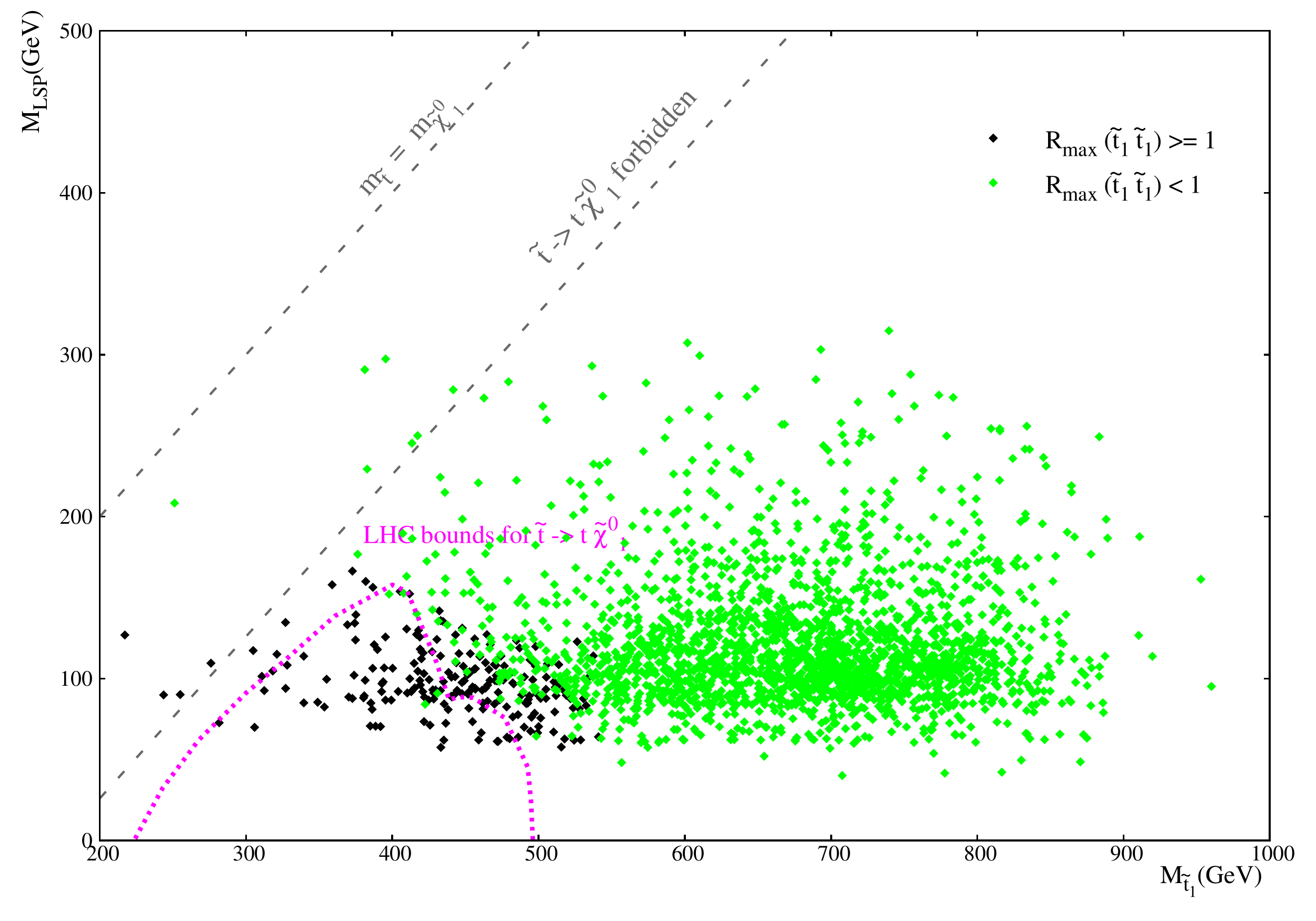}
\includegraphics[width=0.32\columnwidth,angle=270]{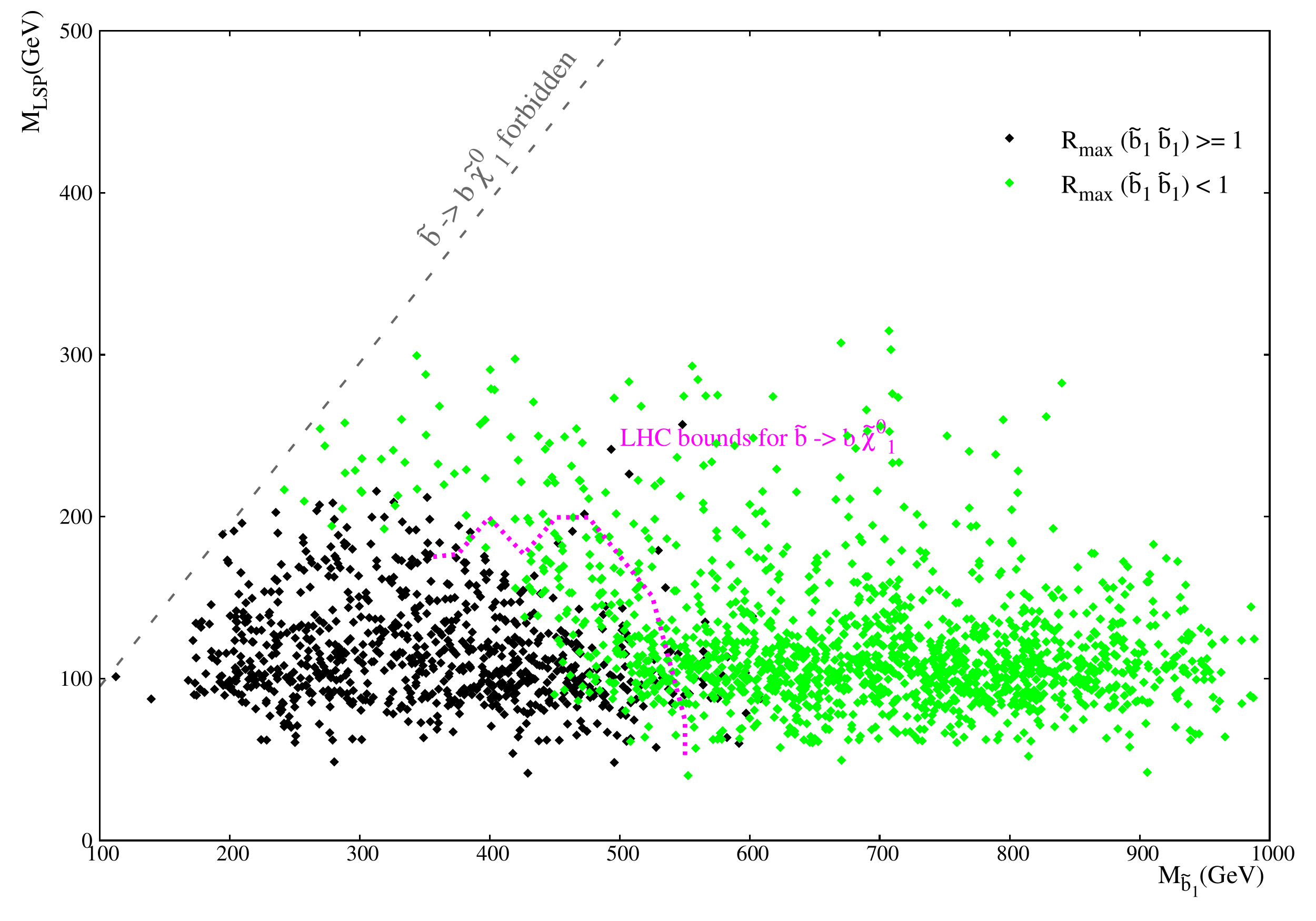}
\caption{The constraints to the solitary signal of $pp\to {\tilde t_1} {\tilde t_1}$ and $pp\to {\tilde b_1} {\tilde b_1}$ are demonstrated, respectively. $R_{\textrm{max}}({\tilde t} {\tilde t})$($R_{\textrm{max}}({\tilde b} {\tilde b})$) means the largest $R$ values in all search channels while the signature only includes the process $pp\to {\tilde t}_1 {\tilde t}_1$ ($pp\to {\tilde b}_1 {\tilde b}_1$).
\label{ct1b1}}
\end{center}
\end{figure}

In Fig. (\ref{ct1b1}), we first examine the constraints on the signal of either the process $p p\to {\tilde t_1} {\tilde t_1}$ or the process $p p\to {\tilde b_1} {\tilde b_1}$ in the $m-m_{\tilde \chi^0_1}$ planes, as shown in the right plot and in the left plot, respectively. The experimental bounds is the same as shown in Fig. (\ref{stop-lsp}). The $R_{\textrm{max}}({\tilde t} {\tilde t})$ or $R_{\textrm{max}}({\tilde b} {\tilde b})$ is obtained by using all kinds of SUSY search analysis approaches (including both $\alpha_T$ and $M_{T2}$, etc.). 

It is interesting to notice that the green points inside the bound curves simple indicate that the branching fraction is too small and yield too small number of signal events to be meaningfully constrained. While the black points outside the bound curves are found to be constrained by other search channels not deliberately designed for either the signal of $pp\to {\tilde t_1}{\tilde t_1}$ or the signal of $pp\to {\tilde b_1}{\tilde b_1}$. For example, the black points outside the bound curve near the point [500, 100] in the $m_{\tilde t}-m_{\tilde \chi^0_1}$ plane and those near the point [600, 100] in the $m_{\tilde b}-m_{\tilde\chi^0_1}$ plane are ruled out by Jets ($M_{T2}$) search channel, while several points near the region of point [500, 250] (with $\tilde b \to b \tilde \chi^0_i$, $\tilde \chi^0_i \to \tilde l l$, and slepton decaying to lepton and LSP) are constrained by the 2SSL search channel. These black points outside the experimental bound curves clearly demonstrate the importance and necessity 
of a comprehensive analysis for a given model.

\begin{figure}[!htb]
\begin{center}
\includegraphics*[width=0.7\columnwidth,angle=270]{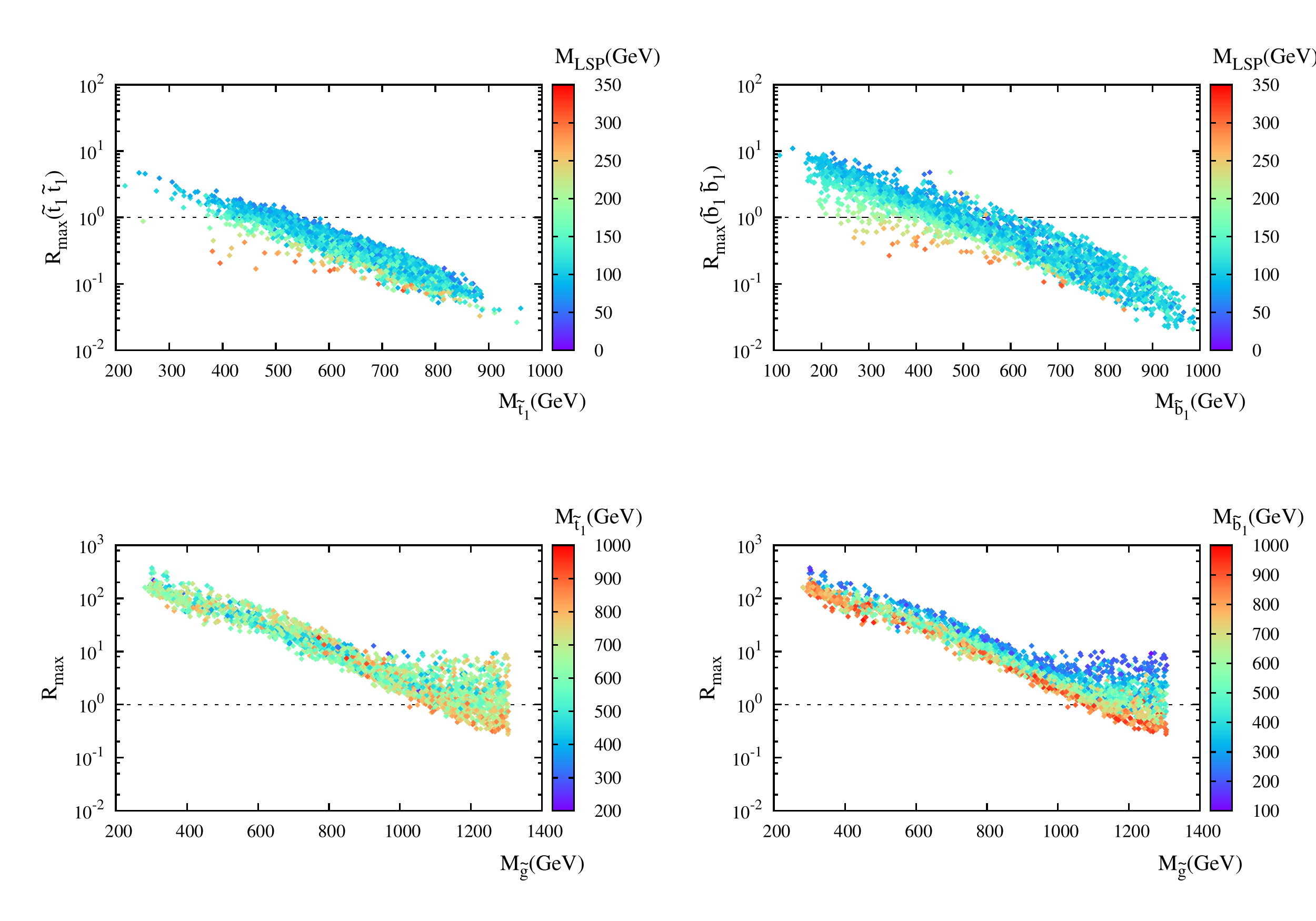}
\caption{In the upper row, the bounds for stop and sbottom are examined where the signals are assumed to be $pp \to {\tilde t_1} {\tilde t_1}$ and $pp \to {\tilde b_1} {\tilde b_1}$, respectively. The y-axis is the $R_{\textrm{max}}$, and the x-axis is the mass of stop and sbottom, respectively.  In the lower row, the bounds for all signals (including $pp \to {\tilde t}_1 {\tilde t}_1$, $pp \to {\tilde b_1} {\tilde b_1}$, and $pp \to {\tilde g} {\tilde g}$, are taken into account. The x-axis is the mass of gluino, the y-axis the $R_{\textrm{max}}$, and the color bar indicates the mass of stop and sbottom, respectively.
\label{ratios}}
\end{center}
\end{figure}

Compared with the results given in \cite{Bi:2011ha,Bi:2012jv}, we observe that when more experimental constraints up to 5 fb$^{-1}$ with $\sqrt{s}=7$ TeV and part of those from the analysis with $\sqrt{s}=8$ TeV are included, the bounds to stop and sbottom have been improved, as clearly demonstrated by the  scattering plots in the upper row of Fig. (\ref{ratios}). It is straightforward to read out that the stop mass can be excluded up to 550 GeV or so, while the sbottom mass can be excluded up to 600 GeV or so.

In Fig. (\ref{ratios}), we show the bounds for the stop, sbottom and gluino. In the lower row, we show bounds against the gluino mass. We observe that most of points have been ruled out/disfavored even if they can be constrained meaningfully neither by the signature of $pp \to {\tilde t}_1 {\tilde t}_1$ nor by the signal of the process $pp \to {\tilde b}_1 {\tilde b}_1$. When gluino is light, say less than 800 GeV, the most stringent bound is from the signatures of $pp\to {\tilde g} {\tilde g}$ due to its large cross section and the bounds show a universal model dependence which is indicated by the width of the band. The width of band when $m_{\tilde g} < 800$ GeV is around 10, i.e. the sensitivity to signature of a given $m_{\tilde g}$ can differ by a factor of 10. In contrast, the sensitivity of signature of $pp \to {\tilde b}_1 {\tilde b}_1$ can differ even larger.

Comparing the left and right plots in the lower row, we observe that the bounds have a strong correlation with the sbottom mass. This correlation, especially at the right corner with 1 TeV $< m_{\tilde g} < 1.3$ TeV where lots of points with a heavy gluino have been excluded, can be attributed to the fact when the cross section $pp\to {\tilde g} {\tilde g}$ is much less than that of $pp\to {\tilde b}_1 {\tilde b}_1$, consequently the real meaningful constraint is actually from $pp\to {\tilde b_1} {\tilde b_1}$. While such a correlation with the stop mass is weak. The width of band near the region 1 TeV $< m_{\tilde g} < 1.3$ TeV become broader, since it is determined by the signature of $pp \to {\tilde b}_1 {\tilde b}_1$, instead of $pp \to {\tilde g} {\tilde g}$.

\begin{figure}[!htb]
\begin{center}
\includegraphics*[width=0.7\columnwidth,angle=270]{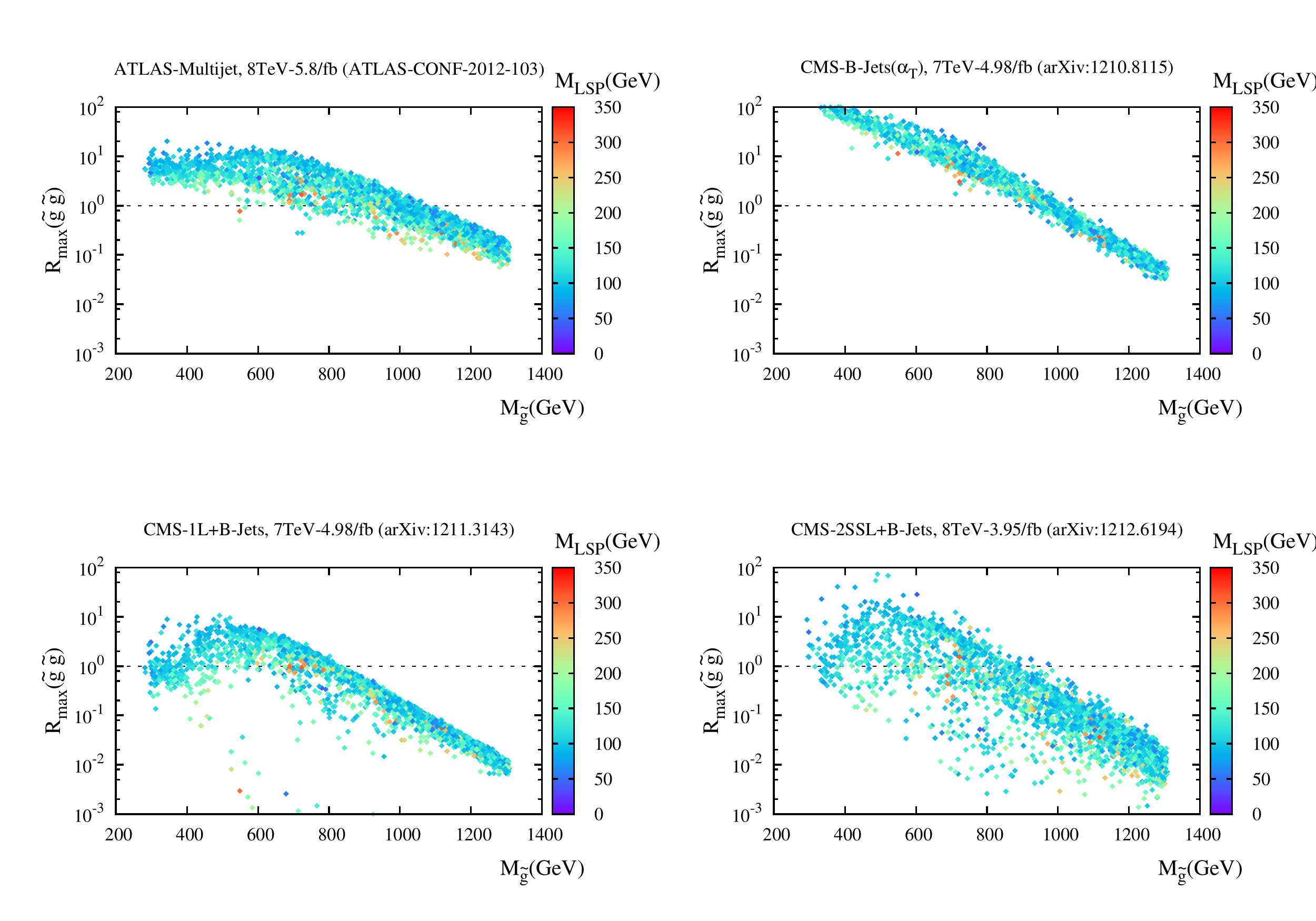}
\caption{Four channels for the signals from $pp\to {\tilde g} {\tilde g}$ are shown. The x-axis is the gluino mass, while the color scale denotes the mass of LSP.
\label{ggratios_lsp}}
\end{center}
\end{figure}

In Fig. (\ref{ggratios_lsp}), we show four representative constraints to the gluino signals from different analysis approaches which are supposed to be sensitive to the signature of the production process $pp\to {\tilde g} {\tilde g}\to t {\tilde t} t {\tilde t}  \sl{E}_T$, for instance. In the upper left plot, the bound from the multijet analysis approach is demonstrated, which can exclude the signals of most models with gluino mass lighter than 600 GeV. In the upper right plot, the bound from the B-Jet plus $\alpha_T$ analysis can exclude the signals below 900 GeV. Meanwhile, this analysis approach enjoys less model dependence than the multijet analysis approach as indicated by the width of the band formed by the points.

In the lower left plot, the bound from one lepton plus B-Jet plus $\sl{E}_T$ is shown and the meaningful constraints can reach up to 800 GeV. At the lower left corner of this plot, there are some points can not be constrained due to the small branching fraction of ${\tilde g} \to t {\bar t} + \sl{E}_T$ and the dominant branching fraction is ${\tilde g} \to b {\bar b} + \sl{E}_T$ or ${\tilde g} \to g + \sl{E}_T$.  While in the lower right plot, the bound from the same sign lepton is demonstrated and the meaningful constraint can reach up to 800 GeV or so. Obviously, both of these two channels rely upon the branching fraction of ${\tilde g} \to t {\bar t} + \sl{E}_T$. Although the same sign lepton mode is clean and has a very tiny SM background, its sensitivity is similar to or worse than the search channel of one-lepton + B jet + ${\sl E}_T$ due to its much smaller branching fraction.

Among these four search channels, it is worthy of remarking that the bounds from the $\alpha_T$ analysis approach with b-tagging is the most stringent and the least model dependent.

\begin{figure}[!htb]
\begin{center}
\includegraphics*[width=0.35\columnwidth,angle=270]{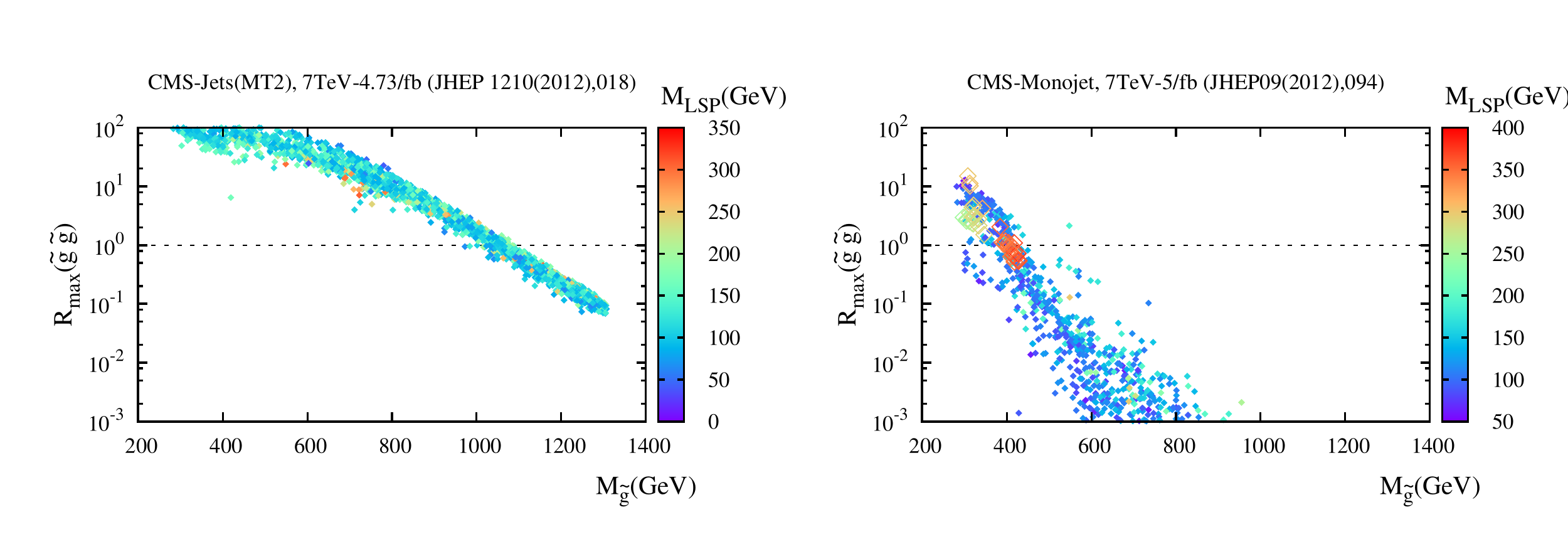}
\caption{The constraints from the $M_{T2}$ analysis approach and the mono-jet channel are demonstrated, respectively. In the right plot, extra points  of the gluino-LSP coannihilation are displayed by the markers of an empty diamond.
\label{mt2andmonoj}}
\end{center}
\end{figure}

\begin{figure}[!htb]
\begin{center}
\includegraphics[width=0.45\columnwidth]{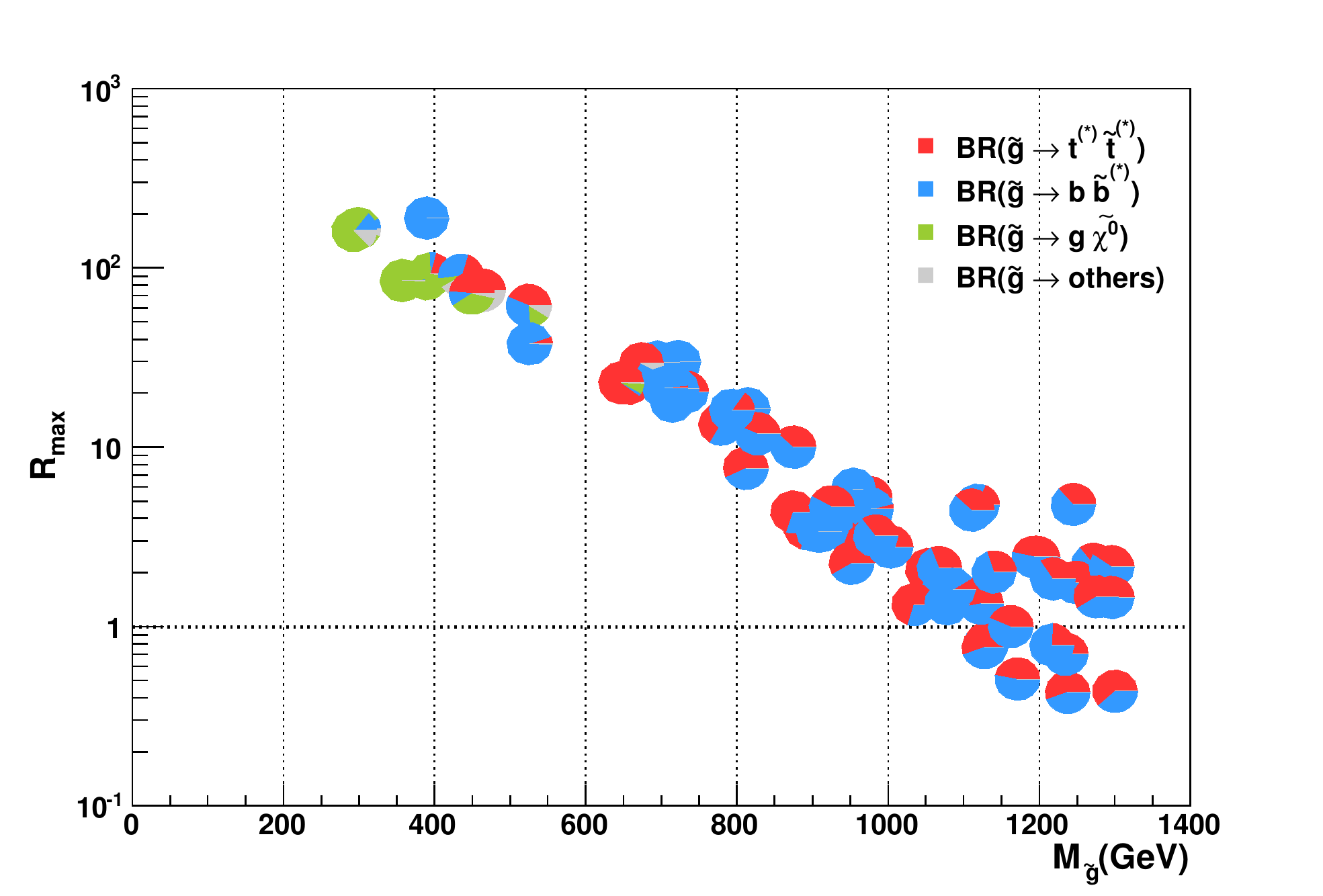}
\includegraphics[width=0.45\columnwidth]{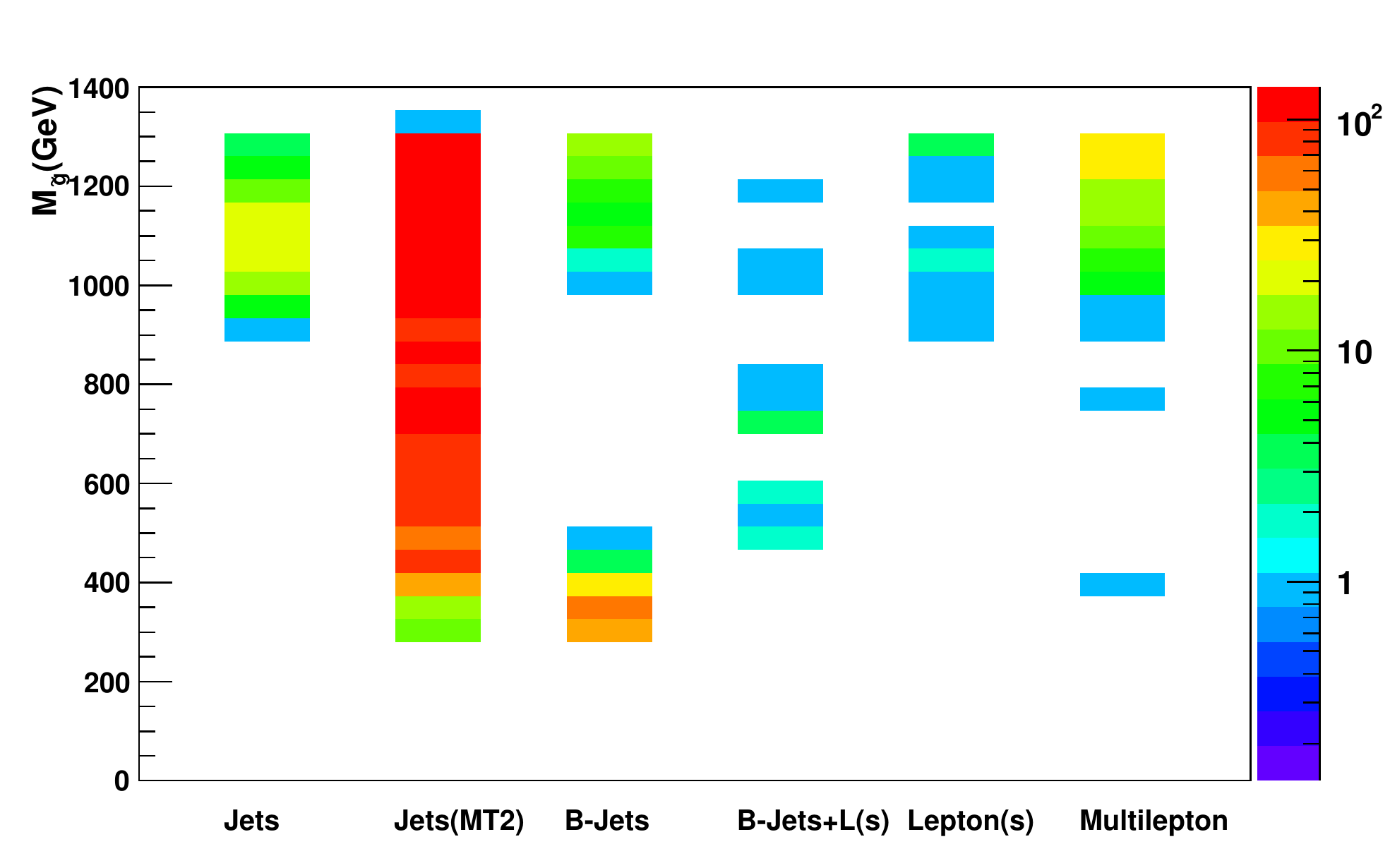}
\caption{At the left plot, the branching fractions dependence is examined by choosing 50 representative points. At the right plot, the statistics on how many models are excluded by which channels are shown.
\label{piechart}}
\end{center}
\end{figure}

We also notice that similar to the $\alpha_T$ approach, the $M_{T2}$ approach also enjoy a good sensitivity and model independence, as shown in the left plot of Fig. (\ref{mt2andmonoj}), where both the constraints from the $M_{T2}$ approach and the constraints from the mono-jet plus $\sl{E}_T$ channel are shown. We notice that the $M_{T2}$ approach can achieve a sensitivity better than the $\alpha_T$ approach for most of points, as be revealed by the statistics shown in the right plot of Fig. (\ref{piechart}). 

Meanwhile, the mono-jet search channel alone can probe the gluino mass up to 400 GeV, as shown in the right plot of Fig. (\ref{mt2andmonoj}). To show the constraint of the mono-jet on the gluino-LSP coannihilation scenarios, we deliberately introduce extra 200 points in this plot as denoted by the markers of an empty diamond.

At the left plot of Fig. (\ref{piechart}), 50 representative points with their branching fractions denoted by pie charts are shown to demonstrate the effects of branching fractions of four main decay chains (say ${\tilde g} \to t \bar t {\sl E}_T$, ${\tilde g} \to b \bar b {\sl E}_T$, ${\tilde g} \to g {\sl E}_T$, and others).  For each category of decay modes, we sum over all on-shell or off-shell decay modes into one. For example, ${\tilde g } \to t^{(*)} {\tilde t}^{(*)}$ means we count on either the on-shell or off-shell decay modes of ${\tilde g } \to t {\tilde t}$, ${\tilde g } \to t^* {\tilde t}$, ${\tilde g } \to t {\tilde t}^*$, ${\tilde g } \to t^* {\tilde t}^*$, and sum over all allowed decay modes. While ${\tilde g } \to b {\tilde b}^{(*)}$ means either the on-shell or off-shell decay modes of ${\tilde g } \to b {\tilde b}$ and ${\tilde g } \to b {\tilde b}^*$. 

It is noticed that the most stringent bounds of gluino with mass lower than 1000 GeV do depend upon the branching fractions, as sensible from the width of the band of $R_{\textrm{max}}$, which denotes the difference of sensitivity for a given $m_{\tilde g}$. This difference can change by a factor of 10 when $m_{\tilde g} < 1000$ GeV, which can be attributed to the fact that when the cross section of $pp\to {\tilde g} {\tilde g}$ is large enough and the branching fractions of ${\tilde g} \to b {\bar b} + {\slashed E}_T$ and ${\tilde g} \to t {\bar t} + {\slashed E}_T$ can be large enough, the constraint the $M_{T2}$ analysis method with b-tagging can always have a good performance, though the branching fraction of 
${\tilde g} \to t {\bar t} \sl{E}_T$ does modify the sensitivity to some degree. When $m_{\tilde g} > 1000$ GeV, the dominant signals might from either $pp\to {\tilde b}_1 {\tilde b}_1$ or $pp \to {\tilde t}_1 {\tilde t}_1$, then the model dependence of the bounds increases.

It is also noticed that when gluino is around 300 GeV $\sim$ 500 GeV and dominantly goes to $g + {\slashed E}_T$, due to the large cross section of $p p \to {\tilde g} {\tilde g}$ and the large mass splitting between the gluino and the LSP (say larger than 100 GeV), either $M_{T2}$ or $\alpha_T$ analysis approach can have a remarkable sensitivity to these points.
Due to the fact that we assume that the squarks of the first two generations are heavier than 1.5 TeV, therefore for most of points the branching fraction of ${\tilde g} \to q q^\prime + {\slashed E}_T$ is typically negligible. 

At the right plot of Fig. (\ref{piechart}), among these 2400 points, the statistical information on how many models are excluded by which channels is provided. It is easy to read out that both the $\alpha_T$ and $M_{T2}$ analysis approaches are overwhelmingly sensitive to most of points, while the search channels with lepton(s) (especially the multiletpon channel with the required lepton number larger than $n_\ell \geq 3$, in which case the standard model background is almost vanishing) start to play a role when neutralinos and charginos are light and their production rates are large. It is also remarkable that when gluino is light and the decay mode ${\tilde g } \to b {\bar b} \tilde \chi_1^0$ is dominant (say around 300 $\sim$ 500 GeV), the search channels with b-tagging can be efficient.

\begin{figure}[!htb]
\begin{center}
\includegraphics[width=0.6\columnwidth,angle=270]{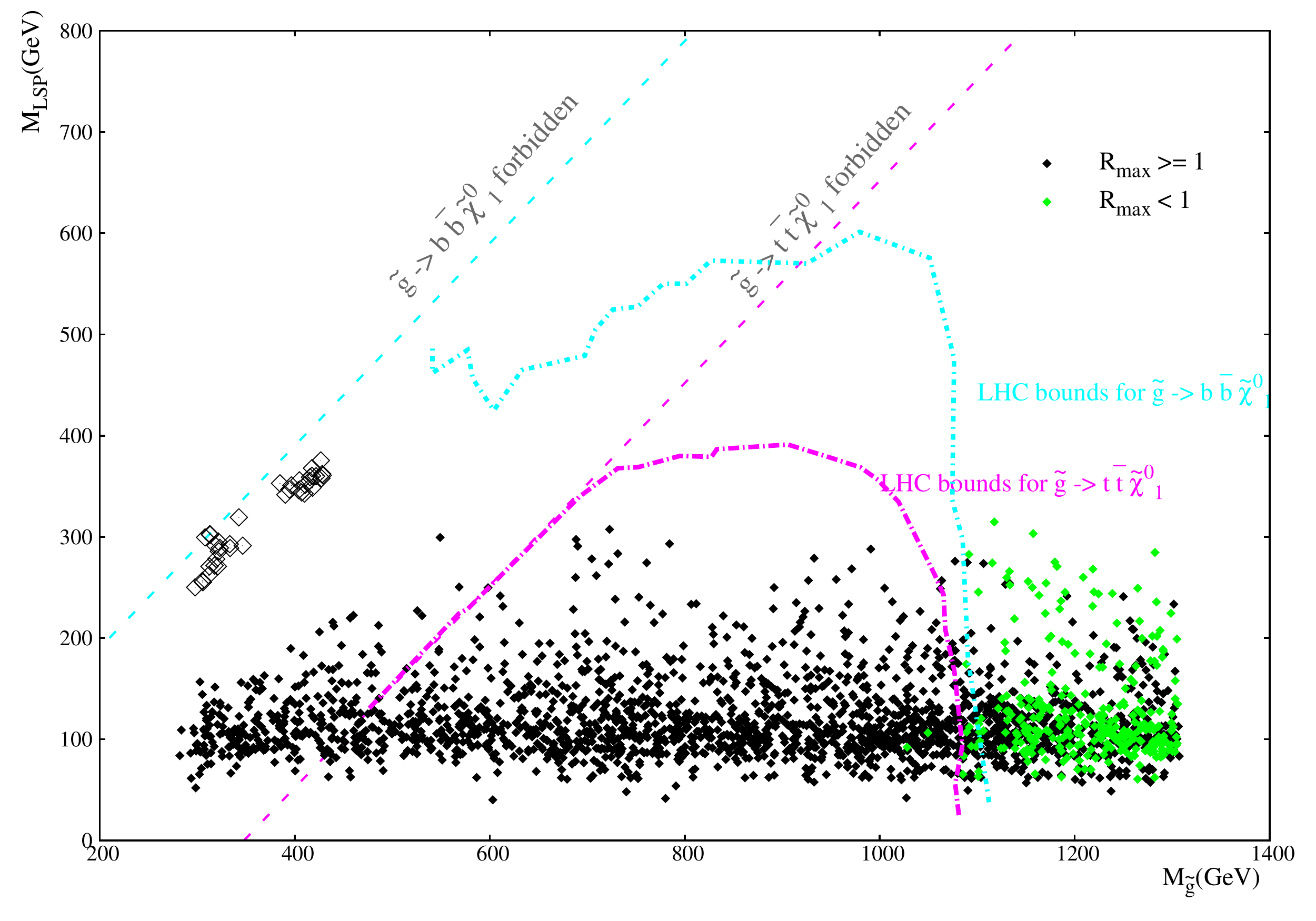}
\caption{The scatter points projected in $m_{\tilde g} - m_{\tilde \chi^0}$ plane are provided, where the black points are heavily disfavored and the green points are safe. The gluino-LSP coannihilation region of points [300,300] and [400, 360] are denoted by markers of an empty diamond.
 \label{mglsp}}
\end{center}
\end{figure}

In Fig. (\ref{mglsp}), we project all points in the $m_{\tilde g} - m_{\tilde \chi^0}$ plane so as to compare with the experimental bounds directly, which have also been shown in Fig. (\ref{gluino-lsp}). There are quite a lots of points outside the experimental bounds are excluded mainly due to the light sbottom in the mass spectra as marked out at Fig. (\ref{ratios}). 

In order to examine those points of near the gluino-LSP coannihilation region, in the first case, we deliberately introduce 200 extra points which are denoted by markers of an empty diamond near the points [300,300] and [400, 360] in Fig. (\ref{mglsp}). It is found that these points can be excluded by both the search channel of mono-jet + ${\sl E}_T$ and the search channel jets + ${\sl E}_T$. 

It is remarkable that for these coannihilation points, the search channel of jets plus $\sl{E}_T$ (say, B-Jets + $\sl{E}_T$, $M_{T2}$ and $\alpha_T$ analysis approaches) has a better sensitivity than the channel of mono-jet plus $\sl{E}_T$ . The underlying reason is that due to the large fraction of boost data sample in the signal events (about $3\sim 6\%$ of the total cross section of $pp\to {\tilde g} {\tilde g}$) when compared with the cross section of the mono-jet events after using the mono-jet search cuts (about $1 \sim 2 \%$ of the total cross section of $pp\to {\tilde g} {\tilde g}$).

For the second case, there is no coannihilation region with mass splitting smaller than 20 GeV. We notice the bound of gluino mass is close to the simplified model due to the large mass splitting $\Delta m = m_{\tilde g} - m_{\tilde \chi^0_1}$ and the energetic final states.  

\section{benchmark points}

We examine benchmark points recommended in literatures. We consider 6 benchmark points labelled by ``NMP'' in Table \ref{6bmp} of NMSSM listed in \cite{King:2012is} and find all of them have been excluded by experiments to a quite high confidence level. We also check two benchmark points labelled as ``EHP'' of NMSSM from \cite{Ellwanger:2012ke} and two benchmark points labelled as ``DET'' of NMSSM from \cite{Das:2013ta}. We also examine the so-called light slepton benchmark point ($\delta M_{\tilde \tau}$) compiled in Table 2 of \cite{Baer:2012yj} with a light slepton sector inspired by anomalous magnet momentum of muon of pMSSM (where both stop and sbottom are also light). For each of benchmark points, we perform the same analysis as those points given in Section III. The bounds obtained from each channel are listed in Table (\ref{6bmp}). 

We present our results in form: $R^{A/C}_{\textrm{search channel, signal region}}$, where R is the maximum ratio of $N_{\textrm{sig}}/N_{\textrm{ul}}$ among all the search channels in each category of signal regions. And the letter ``A" means those from ATLAS collaboration while the letter ``C'' means those from CMS collaborations.

\begin{table}[tbh]
\begin{center}
\scalebox{0.9}{
\begin{tabular}{|c|c|c|c|c|c|c|}\hline
       & jets + ${\sl E}_T$ &B-Jets + ${\sl E}_T$&B-Jets + leptons + ${\sl E}_T$ &
       letpons + jets +  ${\sl E}_T$ & Multilepton(ML)\\ \hline
  NMP1 & $20^{C7}_{M_{T2}}$& $9.7^{C7}_{BJ,\alpha_T}$ &
$7.7^{C8}_{2SSL+BJ}$ & $8.1^{C7}_{1L(tmp)}$ & $1.4^{C7}_{ML}$   
 \\ \hline  
  NMP2 & $20^{C7}_{M_{T2}}$ & $11^{C7}_{BJ,\alpha_T}$ &
$8.3^{C8}_{2SSL+BJ}$ & $6.8^{A8}_{2SSL}$ &  $1.4^{C7}_{ML}$\\
\hline
  NMP3 & $24^{C7}_{M_{T2}}$ & $12^{C7}_{BJ,\alpha_T}$ &
$7.9^{C8}_{2SSL+BJ}$ & $9.2^{C7}_{1L(tmp)}$ &$1.7^{C7}_{ML}$ 
    \\ \hline
  NMP4 & $23^{C7}_{M_{T2}}$ & $12^{C7}_{BJ,\alpha_T}$ &
$6.6^{C8}_{2SSL+BJ}$ & $12^{C7}_{1L(tmp)}$ &  $2.0^{C7}_{ML}$
   \\ \hline
  NMP5 & $22^{C7}_{M_{T2}}$ & $12^{C7}_{BJ,\alpha_T}$ &
$7.2^{C8}_{2SSL+BJ}$ & $11^{C7}_{1L(tmp)}$ &  $1.8^{C7}_{ML}$
     \\ \hline
  NMP6 &  $6.3^{C7}_{M_{T2}}$ & $2.4^{C7}_{BJ,\alpha_T}$ &
$2.2^{A8}_{1L+BJ}$ & $4.3^{C7}_{1L+Jets}$ &  $0.47^{C7}_{ML}$
   \\ \hline \hline
  EHP1 &  $2.6^{C7}_{M_{T2}}$ & $2.2^{C7}_{BJ,\alpha_T}$ &
$0.4^{A8}_{1L+BJ}$ & $1.3^{C7}_{2SSL+Jets}$ &  $0.48^{C7}_{ML}$ 
   \\ \hline
  EHP2&  $2.0^{C7}_{M_{T2}}$ & $1.4^{C7}_{BJ,\alpha_T}$ &
$0.57^{A8}_{1L+BJ}$ & $1.1^{C7}_{1L+Jets(tmp)}$ & 
$0.57^{C7}_{ML}$   \\ \hline \hline
  DET1 & $0.46^{C7}_{M_{T2}}$ & $0.34^{C7}_{BJ, \alpha_T}$ & $0.04^{C8}_{2SSL+BJ}$
& $0.21^{C7}_{1L(tmp)}$ & $0.33^{C7}_{ML}$ \\ \hline
  DET2 & $0.91^{C7}_{M_{T2}}$ & $0.52^{C7}_{BJ, \alpha_T}$ & $0.19^{C8}_{2SSL+BJ}$
& $0.54^{C7}_{1L(tmp)}$ & $0.39^{C7}_{ML}$ \\ \hline  \hline
  $\delta M _{\tilde \tau}$ & $1.2^{A7}_{0L+Jets}$  &  $3.1^{C7}_{BJ,\alpha_T}$ &
  $0.69^{C8}_{2SSL+BJ}$  &  $1.7^{A7}_{2L+Jets}$ &
  $9.96^{C7}_{ML}$    \\ \hline
  \end{tabular}}
  \caption{The maximal ratio of $r=N/N_{exp}$ in each category of all channels are shown here, where $N$ denotes the number of events after all cuts and $N_{exp}$ denotes the allowed number of events by experiments. Numbers in brackets are $CL_s$ for the signal. Super- and sub-scripts denote the LHC collaboration (A means ATLAS and C means CMS), the collision energy ($\sqrt{s}=7$ TeV and $\sqrt{s}=8$ TeV) and the search channel, respectively. The label "tmp" appeared in the 5th column denotes the results obtained by the template approach used by the CMS collaboration \cite{Chatrchyan:2012sca}. }
  \label{6bmp}
 \end{center}
\end{table}

For the original benchmark points labelled as ``NMP" and proposed in \cite{King:2012is}, we observed that the search for the first two generation squarks already rule out all of them. So we modify the mass of the first two generation squarks to 1.5 TeV. It is noticed that the first six benchmark points are very similar in spectra and decays. Each of these six benchmark points has a gluino mass around 700$\sim$800 GeV with electroweakinos lighter than $\sim$500 GeV. It is noticed that for all these six benchmark points, gluino dominantly decays to $\tilde t_1 t$. Therefore there is no surprise that the search channels, like B-Jet(s) + ${\sl E}_T$, lepton + B-Jet + jets +  ${\sl E}_T$, and same sign lepton + B-Jet + jets +  ${\sl E}_T$, are sensitive to the signals of these points. Due to the large mass splitting $\Delta m = m_{\tilde g} - m_{\tilde \chi^0_1}$, the gluino-LSP coannihilation can not occur, which consequently yields energetic visible final states when gluino goes to LSP and consequently leads 
stringent constraints for all these benchmark points. 

As observed in \cite{Bi:2012jv}, these benchmarks points can survive the light stop search bounds when only part of direct search bounds are applied. When more direct search bounds are applied, to save these six benchmark points, it is possible by assuming that gluino mass is higher than 1.5 TeV, as done in \cite{Bi:2012jv}. Except glunio mass, for the third benchmark point, the stop mass must be higher (say 500 GeV, for instance).

For two benchmark points labelled as ``EHP" and proposed from \cite{Ellwanger:2012ke}, it is observed that the most stringent bound is obtained from the search channels jets+${\sl E}_T$, B-Jets + ${\sl E}_T$ (with $\alpha_T$ and $M_{T2}$ analysis approaches), and lepton + jets + ${\sl E}_T$, which are sensitive to the signal from the  production process $pp \to \tilde t_1 \tilde t_1$. These two benchmark points have been studied in \cite{Bi:2012jv} and have been found to be marginally safe when only the ATLAS analysis with 2B-Jets + ${\sl E}_T$ by using 2.05 fb$^{-1}$ dataset and the CMS analysis with 2B-Jets + ${\sl E}_T$ by using 4.98 fb$^{-1}$ dataset with $\sqrt{s}=7$ TeV are applied. Now with more results from the ATLAS and CMS analysis, these two benchmark points start to be in trouble. These two benchmark points can only become safe when the top squark mass is lift up to higher than 600 GeV.

About the two benchmark points labelled as ``DET" and proposed recently in \cite{Das:2013ta} are still marginally safe. The dominant signals of these two benchmark points are from the process $p p \to {\tilde t} {\tilde t}$, though the signals from electroweakino's pair production are also considerable. It is observed that the $M_{T2}$ analysis method can put even more stringent bounds that other analysis method. When the updated SUSY bounds are taken into account, only the first point is marginally safe.

About the light slepton benchmark points labelled as ``$\delta M_{\tilde \tau}$" and compiled in \cite{Baer:2012yj}, although the gluino is heavy, this point has been excluded by the LHC experiments. The most sensitive channel is from the multilepton search mode due to the large cross sections of slepton and electroweakinos' pair production, other search channels sensitive to colored objects also disfavor this point due to its light stop and sbottom.

Based on our analysis, we propose four benchmark points tabulated in Table (\ref{bmp4}) which are safe and can be examined for the future LHC runs.
\begin{table}[tbh]
 \begin{center}
  \begin{tabular}{rl}
  \begin{tabular}[c]{|c|c|c|c|c|}\hline
  Points   &   I   &   II  &  III  & IV\\ \hline  \hline
 $\lambda$    &    0.648   &0.673 &   0.349     & 0.499     \\
 $\kappa$      &    0.323 & 0.252   &   0.415     & 0.140     \\
 $\tan \beta$  &    2.71 & 2.68 &   22         & 9.8       \\
 $\mu_\rm {eff}$ &      303   & 509 &    116       & 213       \\
 $A_\lambda$&     641     & 1208 &   2626      & 2261        \\
 $A_\kappa$ &     -362    & -231  &  -446      & -110        \\
 $M_{\tilde Q_3}$ &     970      & 938 &  854        & 811        \\
 $M_{\tilde U_3}$  &    808      & 620  & 980        & 964        \\
 $M_{\tilde D_3}$  &    275      & 763 &  820        & 957        \\
 $A_t$           &   1792  & 1450 &  1745      & 1833      \\
 $A_b$           &   -60.6 & -2903 &  2887      & 2910       \\
 $M_1$         &    854    & 530 &  786       & 522        \\
 $M_2$         &    964    & 269 &  493       & 260        \\
 $M_3$          &    1094  & 1013 &  1155      & 1174       \\
 $M_{\tilde L}$          &    380   & 371 &  212       & 322        \\
 \hline \hline             
 $m_{H_1}$ &    {\bf 125.7}     & {\bf 125.8} &  118       & 95          \\
 $m_{H_2}$  &   191        & 335 &  {\bf 126.2} & {\bf 126.3}       \\
 $m_{H_3}$  &    827       & 1431 &  2625       & 2186         \\
 $m_{A_1}$  &    424       & 392 &  421       & 149         \\
 $m_{A_2}$  &    824       & 1429 &  2625       & 2186         \\
 $m_{H^\pm}$ & 819         & 1425 &  2625       & 2184         \\ \hline
 $m_{\tilde g}$ &   1209      & 1130   &  1267       &  1285    \\ 
 $m_{\tilde \chi^0_1}$ &  260   & 260   &  99         &  98      \\ 
 $m_{\tilde \chi^0_2}$ &   -324 & 381   &  -129   &  208     \\  \hline

   \end{tabular}                                        
                                                        
   \begin{tabular}[c]{|c|c|c|c|c|}\hline 
  Points   &   I   &   II  &  III & IV \\ \hline  \hline
 $m_{\tilde \chi^0_3}$ &    358 & 496   &  295    &  -235    \\ 
 $m_{\tilde \chi^0_4}$ &    841 & -523   &  518   & 317     \\ 
 $m_{\tilde \chi^0_5}$ &    979 & 570   &   775   &  517     \\ 
 $m_{\tilde \chi^{\pm}_1}$ &   299  & 263&  114   &  181     \\                                                 
 $m_{\tilde \chi^{\pm}_2}$ &    979 & 532&  519   &  317      \\ \hline
 $m_{\tilde t_1}$&  746   & 589  & 790            &  741      \\
 $m_{\tilde t_2}$&   1064 & 993    & 1081         &  1063      \\
 $m_{\tilde b_1}$&   332  & 795  &   848          &  834      \\
 $m_{\tilde b_2}$&  996   & 958   &  878          &  991      \\
 $m_{\tilde \nu_L}$&   376     & 367& 202   &  316      \\
 $m_{\tilde e_L/\tilde \mu_L}$&  382 & 373& 217   &  325      \\
 $m_{\tilde e_R/\tilde \mu_R}$&  381 & 373& 216   &  325      \\
 $m_{\tilde \tau_1}$&  380     &370  &       206   &  319      \\
 $m_{\tilde \tau_2}$&  383     &376   &      227   &  331      \\
\hline  \hline                                      
$BR(B^+ \to \tau^+ \nu_{\tau}) \times 10^4$&1.32  & 1.32 & 1.31  &  1.32    \\ 
 $BR(B \to X_s \gamma) \times 10^4$ &3.63   & 3.40   &  3.89  &  3.56    \\                                                               
$BR(B_s \to \mu^+ \mu^-) \times 10^9$& 3.68   & 3.68   &   3.69  &  3.68    \\ 
$\Omega h^2 $    &    0.006 &   0.003  &     0.01     &      0.10            \\
$\sigma_p^{SI}~(pb) \times 10^9$& 26.6 & 31.2 & 3.2 & 1.1 \\ 
$R^H_{\gamma \gamma}$ &   1.06  & 1.02&  1.25    &  1.14            \\
$R^H_{VV}$ &       1.03    & 1.0  &      0.98     &     1.05            \\
\hline
$BR(\tilde g \to \tilde t_1 t)(\%)$ & 34  & 65  & 46  & 53 \\ 
$BR(\tilde g \to \tilde b_1 b)(\%)$ & 59  & 27  & 28 & 30 \\
$BR(\tilde g \to \tilde b_2 b)(\%)$ & 7  & 8  & 26 & 15 \\\hline
  \end{tabular}    
                                                    
  \end{tabular}
  \label{bmp4}  \caption{Benchmark points for future LHC runs are tabulated.}
 \end{center}
\end{table}

For each case, two benchmark points are presented (mass of LHC Higgs is shown in boldface). For the first case, benchmark point I has a very light sbottom $\sim 300 \rm {GeV}$, which survives from the constraints of the SUSY direct search owing to a heavy LSP(260 GeV). Such a benchmark point might be probed by the full dataset collected with $\sqrt{s}=8$ TeV, as shown in \cite{Yu:2012kj}. Benchmark point II has a relatively light stop (589 GeV), which dominantly decay to $\tilde \chi_2^0 ~t~(29\%)$ and $\tilde \chi_2^+ ~b~(61\%)$ with $\tilde \chi_2^0 \to \tilde \chi_1^\pm ~W^\mp(90\%)$ and $\tilde \chi_2^\pm \to \tilde \chi_1^\pm ~Z~(32\%),~\tilde \chi_1^\pm ~H_1~(23\%),~\tilde \chi_1^0 ~W^\pm~(26\%),~\tilde \chi_2^0 ~W^\pm~(13\%)$. Longer cascade decay chains result in more objects in the final state, many of which are too soft to be reconstructed by detectors of the LHC. 

For the second case, benchmark point III and IV have relatively heavier stops and sbottoms which are safe to current LHC constraints. However, point IV have all five neutralinos and two charginos very light (wino masses are only 317 GeV). It would be expected that the neutralino-chargino search(mainly through tripleton and same-signed leptons signal) at LHC should be sensitive to this point. And this point gets correct relic density by virtue of the large singlino component($\sim 64\%$) of LSP, while the other three points has LSP mostly Higgsino-like (I and III) or Wino-like (II).

It is noticed that masses of the gluinos are around $\sim 1100 - 1300 \rm{GeV}$. The main branching fractions of dominant decay modes of gluino are tabluated. As one can see, the gluino will dominantly decay to $\tilde t_1 t$ and $\tilde b_{1/2} b$ which are exactly the representive simplified models explored by ATLAS and CMS. These benchmark points can be detected in the future LHC runs. And the recent SUSY search results with $\sqrt{s}=8$ TeV presented in the Moroind 2013 EW could be sensitive to these points.   

Another interesting fact is that all the sleptons are lighter than $\sim 500~\rm {GeV}$, currently LHC results can not impose meaningful constraint to slepton sectors (The LHC results can probe the sletpons with masses up to 200 GeV). The future LHC runs at higher collision energies, either $\sqrt{s}=13$ or $\sqrt{s}=14$ TeV can start to probe them.


\section{Discussions and Conclusions}

\begin{figure}[!htb]
    \begin{center}
     \includegraphics[width=0.5\textwidth]{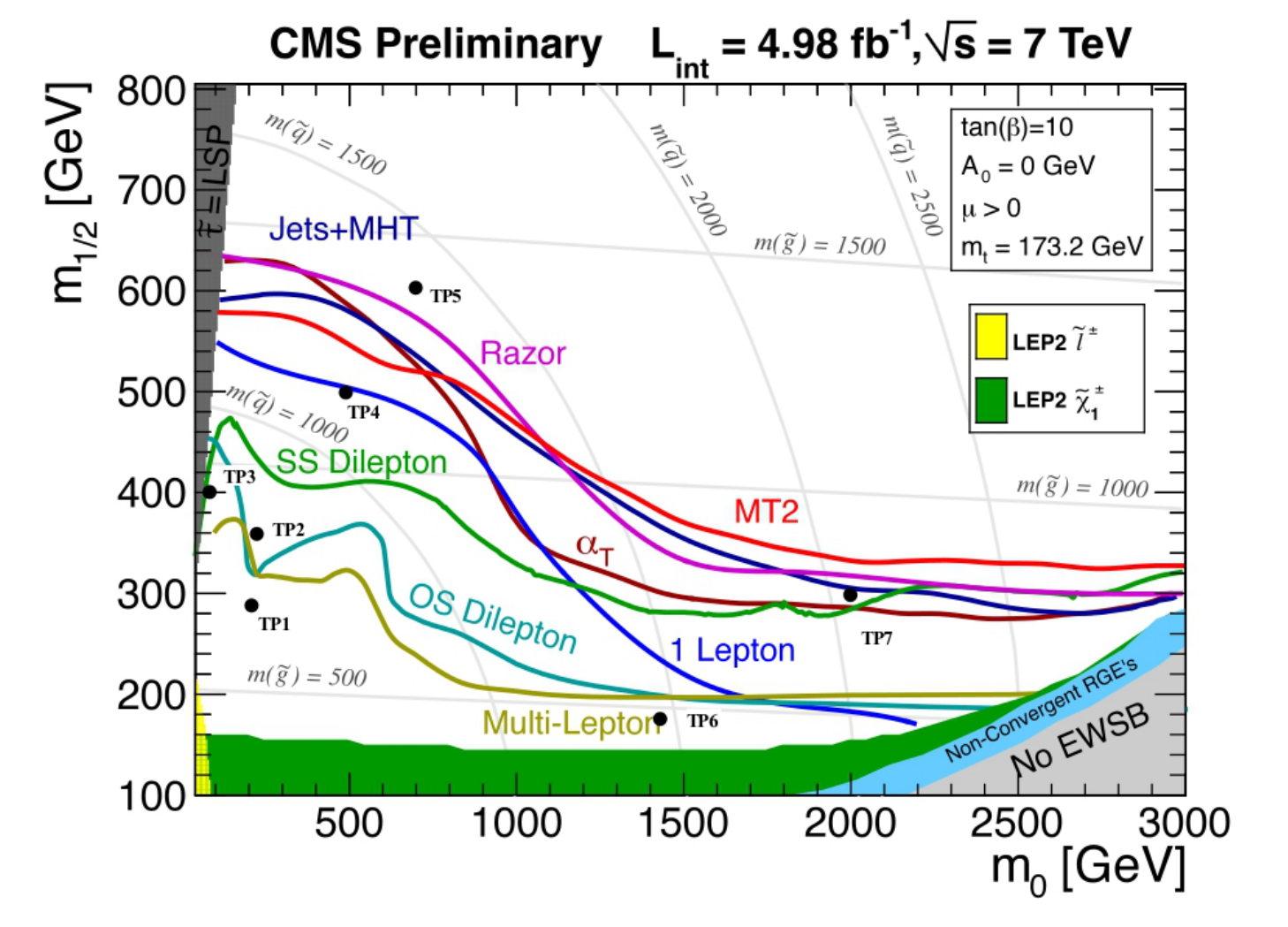}
      \includegraphics[width=0.45\textwidth]{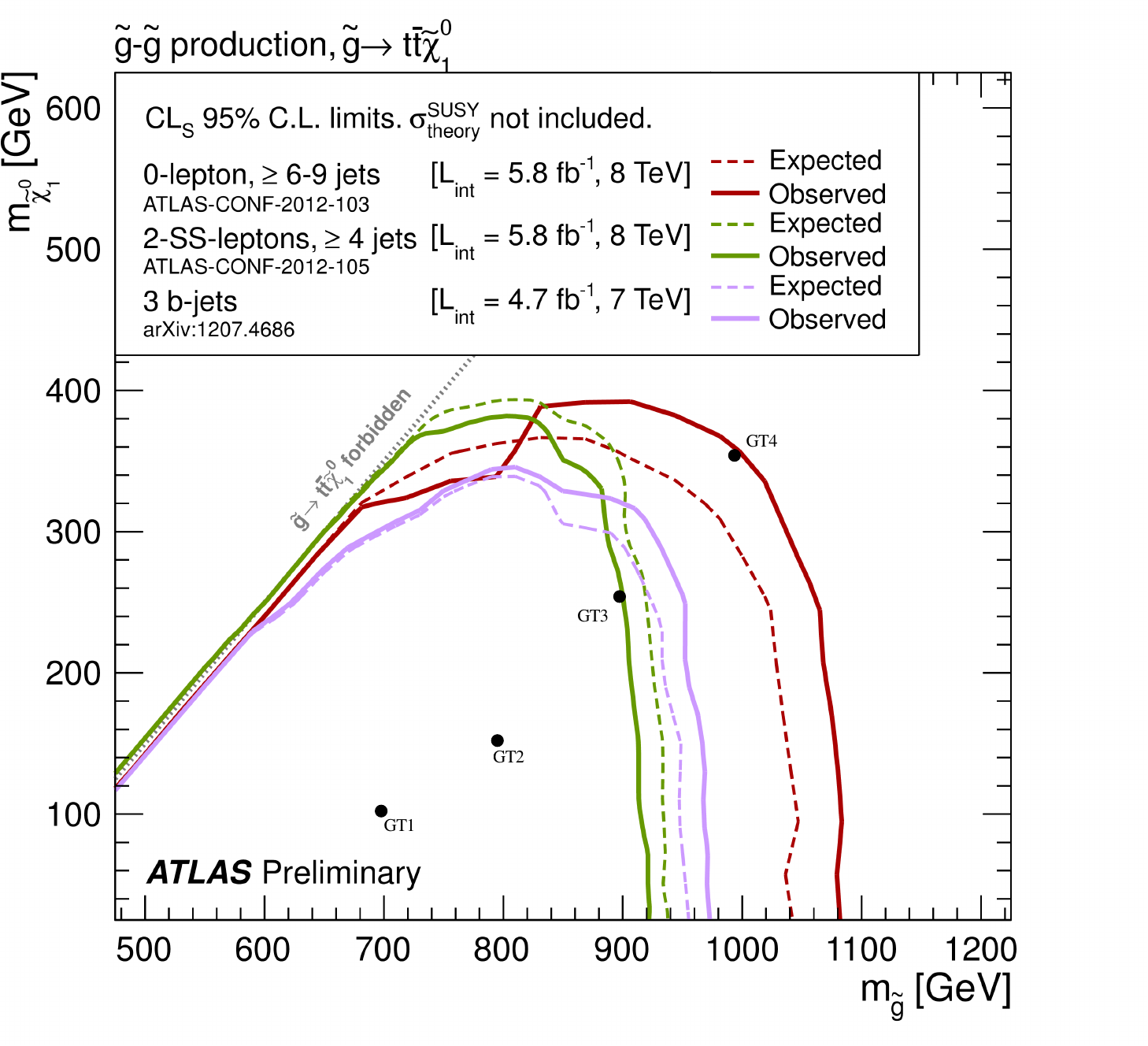}
     \caption{Test points labelled as TP1-TP7 of the CMSSM in the $m_0-m_{1/2}$ plane and those labelled as GT1-GT4 in the simplified models are displayed in our comparison with experimental results.}
     \label{cmssm}
    \end{center}
   \end{figure}

We have extended the analysis \cite{Bi:2011ha,Bi:2012jv} by including more LHC experimental bounds. In order to test the reliability of the results, we perform a thorough check to our package. We have used 21 test points in total to compare with experimental results. These 21 test points have been tabulated in Table (\ref{testp}), where TP1-7 denote benchmark points of the CMSSM as shown in Fig. \ref{cmssm}. And GBs and GTs are benchmark points of the simplified model, with $\tilde g \to t \bar t \tilde \chi_1^0$ or $\tilde g \to b \bar b \tilde \chi_1^0$ with a branching fraction $100\%$. In addition, the test points labelled as "EW1-EW7" denote specific benchmark points designed for the electroweakinos search \cite{Chatrchyan:2012pka}, whose decay modes are also displayed. For test points ``EW1-EW5", we assume $m_{\tilde l}=0.5m_{\tilde \chi_1^{\pm}}+0.5m_{\tilde \chi_1^0}$. 
 \begin{table}[htb]
   \begin{center}
    \scalebox{0.9}{
    \begin{tabular}{|c|c|c|c||c|c|c|c||c|c|c|c|}\hline 
       &  $M_0$  & $M_{1/2}$ &  &  &  $M_{\tilde \chi_2^0/\tilde \chi_1^\pm}$ & $M_{\tilde \chi_1^0}$ & & &  $M_{\tilde g}$ & $M_{\tilde \chi_1^0}$ &\\ \hline 
    TP1&  210  & 285 & CMSSM, LM1  & EW1 & 400 & 200 & \multirow{3}{*}{\begin{minipage}{1.3in}$\tilde \chi_2^0\to l \tilde l (BF=0.5)$, $\tilde \chi_1^{\pm}\to l\tilde \nu, \nu \tilde l$\end{minipage}                                                                                                                                                                   } & GB1 & 800 & 300 & \multirow{3}{*}{$\tilde g \to b \bar b \tilde \chi_1^0$} \\ 
    TP2&   230 & 360 & CMSSM, LM5    & EW2 & 500 & 300 &                                                                                                & GB2 & 1000& 400 &  \\   
    TP3&  85    &  400   & CMSSM, LM6& EW3 & 400 & 250 &                                                                                                & GB3 & 1050 &550 & \\ 
     \cline{8-8} \cline{12-12}
    TP4&  500   &  500   &           & EW4 & 400 & 100 &                                                           $\tilde \chi_2^0\to l \tilde l (BF=1)$,  & GT1 & 700 & 100 & \multirow{4}{*}{$\tilde g \to t \bar t \tilde \chi_1^0$} \\ 
    TP5&  700   &  600   &           & EW5 & 350 & 150 & $\tilde \chi_1^{\pm}\to \nu_{\tau} \tilde \tau_R$                                                                                               & GT2 & 800 & 150 & \\ 
     \cline{8-8} 
    TP6&  1450  &  175   & CMSSM, LM9& EW6 & 200 & 75  & $\tilde \chi_2^0\to Z \tilde \chi_1^0$,              & GT3 & 900 & 250  &\\ 
    TP7&  2000  &  300   &      & EW7 & 200 & 50  & $\tilde \chi_1^{\pm}\to W^{\pm} \tilde \chi_1^0 $                                                                                                 & GT4 & 1000& 350  &\\ \hline
    \end{tabular}} 
    \label{testp}
    \caption{All test points are tabulated.}                                                                                                                                        
   \end{center}
  \end{table}

The results of 21 points are tabulated in Table (\ref{chktbl}). In the table, search channels are arranged by the order of the ATLAS and CMS Collaborations and in the order of collider energies, with datasets of $\sqrt{s}=7$ TeV first and followed by the datasets of $\sqrt{s}=8$ TeV. For the multilepton channel, we list the check results separately. 

In each row of the Table (\ref{chktbl}), one can read the results for each search channel. For example, it is noticed that the search channel ``1-2B-Jets + 1-2L" is insensitive to all our test points, since this channel is designed for very light stop (with mass similar to, or lighter than top quark) search. 

From Table (\ref{chktbl}), we can observe a fact that for test points ``TP1-TP7", ``GB1-GB3", and ``GT1-GT4", the $M_{T2}$ observable is quite sensitive to most of signals, similar to our observations in Section III. Furthermore, the results for the ``Multilepton'' search channel also agree with experimental ones \cite{Chatrchyan:2012pka} very well. It is also obvious that the sensitivity of a specific search channel can perform better for the dataset of $\sqrt{s}=8$ TeV than that of $\sqrt{s}=7$ TeV due to the enhancement of cross sections, as demonstrated in the channels ``Multijet" of the ATLAS and ``2SSL + B-\textrm{Jets}" of the CMS.

\begin{figure}[!htb]
    \begin{center}
      \includegraphics[width=0.4\textwidth]{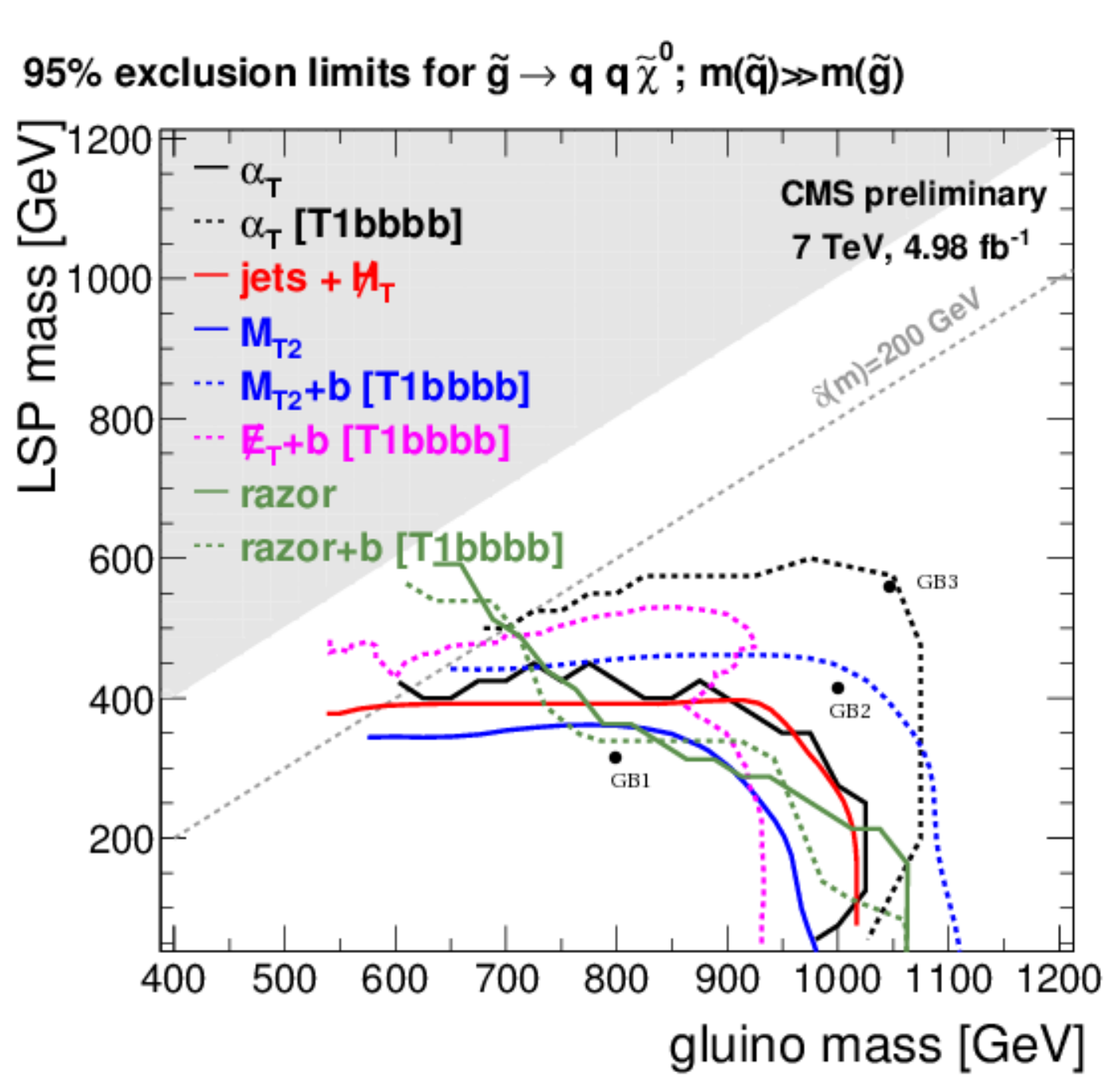}
      \includegraphics[width=0.4\textwidth]{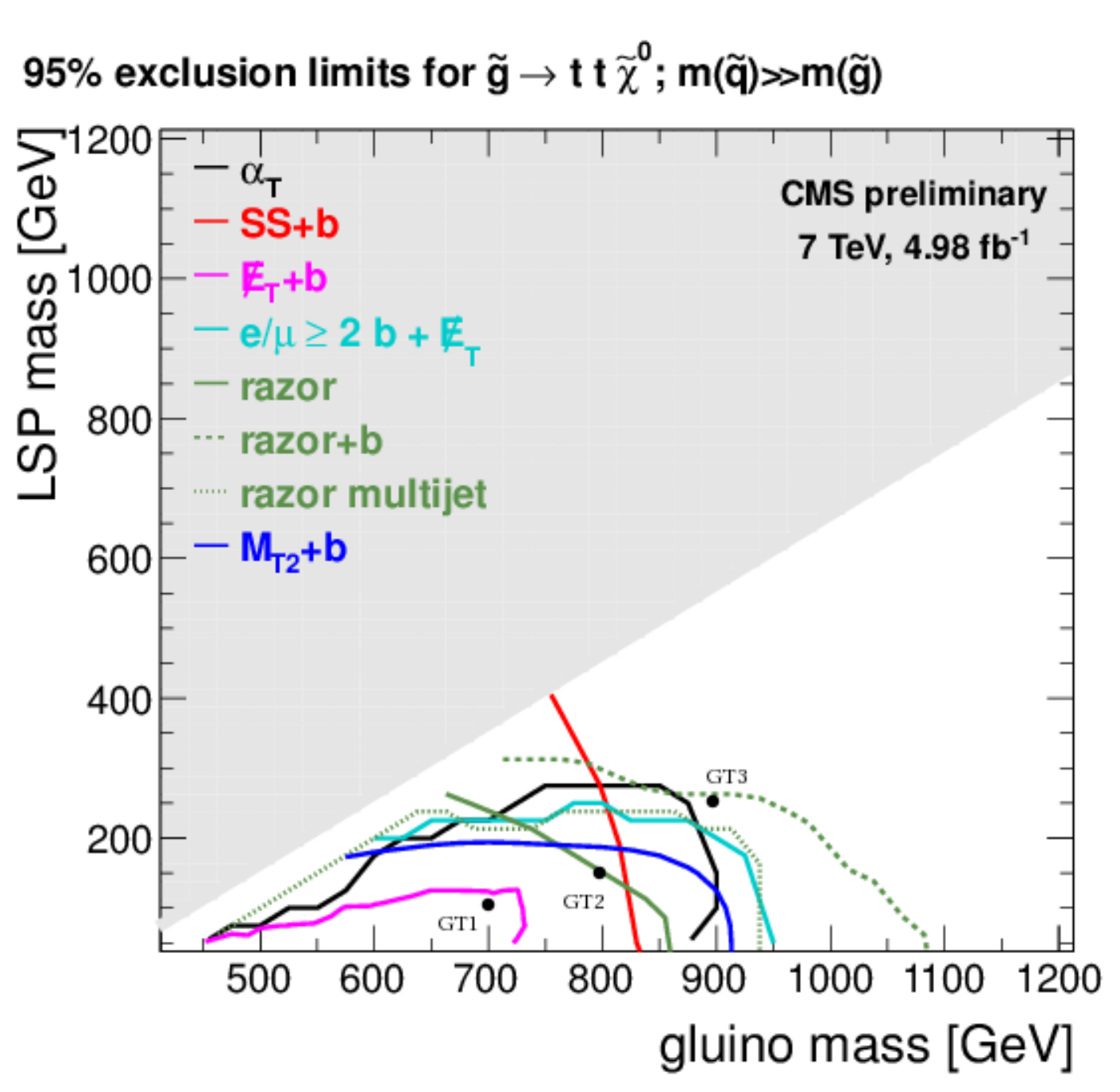}
      \caption{Test points labelled as GB1-GB3 and GT1-GT3 in the simplified models are displayed in our comparison with experimental results.}
      \label{sms}
    \end{center}
   \end{figure}

In the ``Jets+MHT'' search channel, one can read out that for seven CMSSM test points five are all excluded except for TP5 and TP7, which is in agreement with the results of CMS-7TeV as shown in Fig. \ref{cmssm}. While the result for GB1 shows a minor deviation. As shown in Fig. \ref{sms}, this test point should have been excluded by the search channel ``Jets + ${\sl H}_T$", while our result underestimates the constraint with a $R_{\textrm{max}}=0.8$. 

We observed that most of our results agree with experimental results, though there are some results showing deviations. For example, the ``GB3" is under-constrained by the 2SSL + B-Jets channel, and the ``GT4" is over-constrained by the $M_{T2}$ analysis approach and is under-constrained by the multijet channel. Compared with experimental results, these deviations of $R$ values in Table (\ref{chktbl}) can be typically around $\pm 30\%$ or so, which can only affect the results of those points near the exclusion edges of experimental results. For example, the results of ``GB3" 
and ``GT4" should be excluded by the corresponding experimental search channels, but can survive in our analysis. When a point is far away from the edge and inside the exclusion region, our results are trustable, like the test 
point ``GT1". 

Meanwhile, we notice that when the gluino mass is shifted by $\pm 40$ GeV, the deviations can be removed and agreements can be achieved. The deviations in $R$ can be attributed to the small fluctuations from Monte Carlo simulations and the difference between the fast detector simulation and the real detector effects. Our bottom line is that our results at least motivate experimentalists to perform a more detailed and serious bounds when all types of uncertainties are taken into account. 

 \begin{table}[htb]
  \begin{center}
  \scalebox{0.9}{
   \begin{tabular}{|c||c|c|c|c|c|c|c|||c|c|c||c|c|c|c|}\hline
                 & TP1 & TP2 & TP3 & TP4 & TP5 & TP6 & TP7 & GB1 & GB2 & GB3 & GT1 & GT2 & GT3 & GT4 \\ \hline \hline
    Jets     &  21 &  12   & 7.6  &  6.4   & 1.6   & 5.4 & 1.4  & 1.3  & 0.42  & 0.16 & 2.9 & 1.9  & 0.84 & 0.37  \\ \hline
    Multijet &  1.7 &  0.72  & 0.13 & 0.11 & 0.02 & 2.3 & 0.63 & 0.09 & 0.02  & 0.01 & 1.8 & 0.79 & 0.29 & 0.11  \\ \hline
    B-Jets  &   4.1 &  2.2   & 0.39 & 0.24 & 0.04 & 3.6 & 0.56 & 4.8  & 0.86  & 0.45 & 4.8 & 2.2  & 0.91 & 0.37  \\ \hline
    Jets[Heavy Stop]& 17  &  6.1 &  1.8 & 0.59 & 0.09 & 7.4  & 1.1 & 1.8  & 0.33 & 0.18 & 5.9  & 3.2  & 1.4  & 0.63 \\ \hline
    1-2B-Jets+1-2L  &  0  &   0  &  0   & 0    & 0     & 0    & 0   & 0    & 0    & 0    & 0    & 0    & 0    & 0    \\ \hline
    2L+Jets         &0.07 &  0.01& 0.05 & 0& 0     & 0.15 & 0.01& 0    & 0    & 0    & 0.01 & 0    & 0    & 0    \\ \hline
    2L+Jets[Medium Stop]& 1& 0.36& 0.54 & 0.06& 0.01 & 0.28 & 0.09& 0    & 0    & 0    & 0.76 & 0.33 & 0.12 & 0.06 \\ \hline
    1L+Jets         & 3.6 & 0.95 & 0.73 & 0.10 & 0.02 & 1.0  & 0.24& 0    & 0    & 0    & 2.8  & 1.2  & 0.47 & 0.19 \\ \hline
     \hline            
    Multijet& 18  & 7.8   & 5.0   & 1.5  & 0.17  & 1.8  & 0.27 & 0.80 & 0.27 & 0.11 & 0.46 & 0.39 & 0.19 & 0.09 \\ \hline
    $M_{T2}$     & 50  & 26    & 9.5   & 3.9  & 0.68  & 42   & 5.9  & 8.4  & 1.6  & 0.81 & 16   & 6.4  & 2.5  & 1.0  \\ \hline
    B-Jets  & 9   & 5.6   & 2.6   & 0.91 & 0.17  & 2.9  & 0.37 & 3.2  & 0.53 & 0.30 & 2.6  & 1.4  & 0.59 & 0.24 \\ \hline
(B-)Jets, $\alpha_T$& 57  & 25 & 13 & 3.8&0.63   & 30   & 3.1  & 2.5  & 0.71 & 0.25 & 4.7  & 2.2  & 0.84 & 0.31 \\ \hline
    1L+B-Jets & 1.0 & 0.66 &0.31 & 0.08 & 0.01 & 0.58 & 0.37 & 0    & 0    & 0    & 2.6  & 1.2  & 0.45 & 0.18 \\ \hline
    2SSL+B-Jets& 2.4 & 0.64 & 0.94 & 0.07& 0.01 & 1.4  & 0.59 & 0    & 0    & 0    & 4.4  & 1.9  & 0.67 & 0.26 \\ \hline
    1L(tmp)   & 7.1 & 4.6  &2.9   & 0.74 & 0.14  & 2.3  & 1.4  & 0    & 0    & 0    & 5.7  & 3.4  & 1.5  & 0.63 \\ \hline
    2SSL       & 11  & 5.4   & 4.5 & 0.39 & 0.06 & 15   & 1.4  & 0.09 & 0.01 & 0.01 & 7.2  & 2.7  & 1.0  & 0.38 \\ \hline
    2OSL       & 11  & 5.3  & 4.3  & 0.58 & 0.09 & 3.7  & 0.78 & 0.05 & 0.01 & 0.01 & 2.7  & 1.3  & 0.57 & 0.25 \\ \hline
   $\tau$(s)+Jets& 24 & 11   & 6.3  & 1.0& 0.16  & 4.4  & 0.45 & 0.54 & 0.15 & 0.07 & 1.4  & 0.65 & 0.26 & 0.13 \\ \hline
    Z-boson+Jets & 5.9 & 0.35 & 0.94 & 0.05 &0.01& 0.37 & 0.15 & 0    & 0    & 0    & 0.27 & 0.11 & 0.04 & 0.02 \\ \hline
    Monojet      & 4.2 &  1.4 & 1.2  & 0.12 & 0.02&0.03& 0.01& 0.02 & 0    & 0    & 0    & 0    & 0    & 0    \\ \hline \hline
    & EW1 & EW2 & EW3 & EW4 & EW5 & EW6 & EW7 & -- & -- & -- & -- & -- & -- & -- \\ \hline 
    Multilepton  & 2.0 &  0.61 & 0.81 & 1.5 &1.2 & 0.61 &0.77 & --&-- & --&-- & -- & -- & -- \\ \hline
    \hline \hline \hline \hline
    Multijet    & -- & -- & -- & -- & -- & -- & -- & -- & -- & -- & 2.8 & 2.2 & 1.2  & 0.60\\ \hline
     Jets        & -- & -- & -- & -- & -- & -- & -- & -- & -- & -- & 4.2 & 1.8 & 0.79 & 0.32\\ \hline
     2SSL         & -- & -- & -- & -- & -- & -- & -- & -- & -- & -- & 4.4 & 2.7 & 1.3  & 0.61\\ \hline
     1L+Jets     & -- & -- & -- & -- & -- & -- & -- & -- & -- & -- & 5.5 & 2.1 & 0.88 & 0.35\\ \hline
    \hline
     2SSL+B-Jets & -- & -- & -- & -- & -- & -- & -- & -- & -- & -- & 4.9 & 2.4 & 1.1  & 0.42\\ \hline
     
   \end{tabular}}
   \caption{Results for each individual search channels are presented for checking. Numbers in the table are defined as $R=N_{sig}/N_{UL}$, where $R>1$ means excluded by experiments. The slots with a hyphen mean we haven't look at these channels since the signal is expected to vanishing. }
   \label{chktbl}
  \end{center}
 \end{table}

Based on our analysis shown above, below we comment on light gluino models in literatures. Models in the coannihilation region as demonstrated by LG3-5 given in \cite{Chen:2010kq} can be ruled out due to the mono-Jet search and by the $\alpha_T$ and $M_{T2}$ analysis (due to the boosted sample events) from the CMS if the branching fraction ${\tilde g} \to g \tilde \chi^0_1$ is dominant. Due to the large mass splitting between the gluino and the LSP and the energetic final states, light gluino scenarios with mass range [500,700] GeV in the so-called Higgs pole region of the minimal universal SUGRA \cite{Feldman:2011me} is heavily disfavored.
The two benchmark points presented in Ref.~\cite{Ajaib:2010ne} has been excluded due to their large cross sections and large mass splittings. We also find that even for the coannihilation region (say $m_{\tilde g } - m_{\tilde \chi^0} < 20$ GeV) the mono-jet bounds can be valid for gluino mass up to 400 GeV at least, as shown by the right plot of Fig. (\ref{mt2andmonoj}).

The last but not the least, in our scanning, we have not found points near the region $m_{\tilde g} \approx m_{\tilde \chi^0} \sim 500$ GeV.  Such models should exist since there is no upper bounds for the LSP, as demonstrated in the Fig. (8c) of  \cite{Gherghetta:2012gb} by few red points, where the scanning ranges for parameters of neutralino and gluino are wider than those in our scanning.

With results shown above, we would like to conclude: 1) the parameter space of the natural NMSSM has been significantly shrunken by the direct SUSY search bounds from the ATLAS and CMS collaborations; 2) The gluino-LSP coannihilation region in the first case of interpretations of the Higgs boson data has been completely ruled out by the boosted signal samples. We can safely claim that the lightest gluino mass must be larger than 400 GeV for the gluino-LSP coannhilation region.

\noindent{\bf Note:} When this work is finished, in the updated results presented at the Rencontres de Moroind 2013 EW, we notice that the gluino mass must be heavier than 1.3 TeV when the squarks of the third generation are assumed to be light. Experimental bounds nearly approach the upper limit of gluino mass of the natural SUSY.

\noindent {\bf Ackownoledgement}\\
We would like to thank Chunli Tong very much for his contribution to the MCMC method at the early stage of this project. We also thank Qiang Yuan from IHEP, CAS for his kind explanation on the advantages of the MCMC method. We are indebted to Zhen-Wei Yang and Yuan-Ning Gao from HEP group of Tsinghua University and Xiao-Rui Lv and Yang-Heng Zheng from UCAS for kind helps on using the PC farms of their groups. 
This research was supported in part
by the Natural Science Foundation of China
under grant numbers 10821504, 11075194, 11135003, 11275246 (TC, JL, TL), 11175251 (QY),
and by the DOE grant DE-FG03-95-Er-40917 (TL).

\end{document}